\UseRawInputEncoding 
\documentclass[fleqn,usenatbib]{mnras}

\usepackage{newtxtext,newtxmath}
\usepackage[T1]{fontenc}
\usepackage{ae,aecompl}

\usepackage{graphicx}	

\usepackage{amsmath}	
\usepackage{amssymb}	
\usepackage[colorinlistoftodos]{todonotes}

\newcommand{\later}[1]{\textcolor{magenta}{To be written later. }}

\newcommand{\mass}{M_\mathrm{200b}}
\newcommand{\mar}{\Gamma_\mathrm{dyn}}
\newcommand{\fmis}{f_\mathrm{mis}}
\newcommand{\rmis}{R_\mathrm{mis}}
\newcommand{\ds}{\Delta\Sigma}
\newcommand{\rockstar}{\fontfamily{lmtt}\selectfont ROCKSTAR}
\newcommand{\sparta}{\fontfamily{lmtt}\selectfont SPARTA}
\newcommand{\scatter}{\sigma_{\Gamma_\mathrm{dyn}|\mathrm{obs}}}
\newcommand{\halotools}{\fontfamily{lmtt}\selectfont Halotools}
\newcommand{\dq}{\fontfamily{lmtt}\selectfont Dark Quest}

\title[Detecting MAR]{Beyond Mass: Detecting Secondary Halo Properties with Galaxy-Galaxy Lensing}

\author[Xhakaj et al.]{Enia Xhakaj$^{1}$\thanks{E-mail: exhakaj@ucsc.edu}, Alexie Leauthaud$^{1}$, Johannes Lange$^{1,2}$,  Andrew Hearin$^{3}$,  
\newauthor
Benedikt Diemer$^{4}$, Neal Dalal$^{5}$
\\
\\
$^{1}$Department of Astronomy and Astrophysics, University of California, Santa Cruz, 1156 High Street, Santa Cruz, CA 95064 USA \\
$^{2}$ Kavli Institute for Particle Astrophysics and Cosmology and Department of Physics, Stanford University, CA 94305, USA \\
$^{3}$High-Energy Physics Division, Argonne National Laboratory, Argonne, IL 60439, USA \\
$^{4}$Department of Astronomy, University of Maryland, College Park, MD 20742, USA\\
$^{5}$Perimeter Institute for Theoretical Physics, 31 Caroline Street N., Waterloo, Ontario, N2L 2Y5, Canada
}

\date{Accepted XXX. Received YYY; in original form ZZZ}

\pubyear{\today}

\begin{document}
\label{firstpage}
\pagerange{\pageref{firstpage}--\pageref{lastpage}}
\maketitle
\begin{abstract}

Secondary halo properties beyond mass, such as the mass accretion rate (MAR), concentration, and the half mass scale, are essential in understanding the formation of large-scale structure and dark matter halos. In this paper, we study the impact of secondary halo properties on the galaxy-galaxy lensing observable, $\Delta\Sigma$. We build an emulator trained on N-body simulations to model $\Delta\Sigma$ and quantify the impact of different secondary parameters on the $\Delta\Sigma$ profile. We focus on the impact of MAR on $\Delta\Sigma$. We show that a 3$\sigma$ detection of variations in MAR at fixed halo mass could be achieved with the Hyper Suprime Cam survey in combination with a proxy for MAR with scatter $\scatter<1.5$. We show that the full radial profile of $\Delta\Sigma$ depends on secondary properties at fixed halo mass. Consequently, an emulator that can perform full shape fitting yields better than  2 times improvement upon the constraints on MAR than only using the outer part of the halo. Finally, we highlight that miscentering and MAR impact the radial profile of $\Delta\Sigma$ in a similar fashion, implying that miscentering and  MAR need to be modeled jointly for unbiased estimates of both effects. We show that present-day lensing data sets have the statistical capability to place constraints on halo MAR. Our analysis opens up new possibilities for observationally measuring the assembly history of the dark matter halos that host galaxies and clusters. 
\end{abstract}




\section{Introduction}\label{intro}

In the concordance cosmological model, dark matter is the dominant form of matter in the Universe. Small density perturbations in the early Universe, coupled with the effects of gravity, lead to the formation of virialized structures, called dark matter halos \citep[e.g.,][]{Gunn1972OnEvolution, Fillmore1984Self-similarUniverse}.
The rate at which halos accrete dark matter can be crucial in understanding the growth of both large-scale structure and galaxies. It has long been known that halo mass accretion rate (MAR) is correlated with the baryon cooling rate \citep[e.g.,][]{whiterees78, fall_80, blumenthal86}. This, in turn, influences the formation of clusters and luminous galaxies. From a galaxy formation perspective, many recent studies have confirmed that MAR correlates with star formation and is therefore valuable in probing galaxy growth \citep[e.g.,][]{diemer13, wetzel_nagai15,  odonnell2020}.

Furthermore, it has been shown that halo clustering depends on secondary halo properties other than mass, the most notable of which is mass accretion history \citep[MAH, e.g.,][]{wechsler2001, gao2005, Wechsler2006, li2008}. This effect is commonly referred to as ``halo assembly bias''. Early work on this topic primarily focused on its manifestation on large scales. Recent work has highlighted an aspect of assembly bias that is especially relevant for the present paper: secondary halo properties beyond mass produce distinct, scale-dependent features upon large-scale structure observables that have the potential to serve as signatures of the particular secondary property \citep{Sunayama_scale_dependence, behroozi_scale_dependence}. Having an adequate understanding of the spatial distribution of halos is essential in interpreting how galaxies cluster. Thus, understanding the MAHs of halos is a key ingredient in accurately modeling the dark matter distribution in the Universe. Finally, MAR can also be used as an independent cosmological probe to constrain $\sigma_8$ and $\Omega_{\rm m}$. For example, \citet{hurier19} showed that certain definitions of MAR can be nearly independent of halo mass. As such, MAR can constrain the $\sigma_8$-$\Omega_{\rm m}$ degeneracy without needing to calibrate a mass-observable relation. 

Given the importance of MAR in understanding the fundamentals of structure formation, a natural question is whether it can be measured in observations. Halo outskirts have shown strong prospects in detecting MARs. For instance, \citet{Diemer2014DependenceRate} model density profiles as comprised of three parts:
an Einasto profile (i.e., the 1-halo term), a power-law term (i.e., the two-halo term), and a  function that models the transition of the one and two halo terms by truncating the Einasto profile. The radius at which this truncation happens is defined as $r_\mathrm{t}$. \citet{Diemer2014DependenceRate} show that a tight correlation $r_t$ and the splashback radius. Both these quantities are, in turn, determined by MAR. The splashback feature has been previously proposed as a potential observable for detecting MARs \citep{Diemer2014DependenceRate, xhakaj19}. It is well known that the splashback feature is tightly correlated with MAR  \citep[e.g.,][]{Diemer2014DependenceRate, Adhikari2014SplashbackHalos, More2015TheMass}. \citet{xhakaj19} studied how well this correlation can be detected with present and future surveys. They showed that detecting MAR through the splashback radius would be feasible, assuming ideal observational proxies of both halo mass and MAR. Thus both $r_t$ and the splashback radius can be used as observational probes to detect MAR.

Moreover, it has long been known that halo profiles correlate with concentration.  \citet{wechsler02} showed that the concentration parameter from the Navarro-Frenk-White profile \citep[NFW,][]{Navarro1996TheHalos} correlates with the halo mass assembly history. \citet{ludlow13} further showed that the Einasto profile shape parameters correlate with concentration and halo assembly history. However, both of these studies use concentration as a proxy for MAH. This is because concentration can be conveniently measured as the ratio of the virial to the scale radius of the halo $(c_{\rm vir} = \frac{R_{\rm vir}}{R_{\rm s}})$, where $R_{\rm s}$ is the scale radius and $R_{\rm vir}$ is the virial radius of the halo. The scale radius $R_s$ is one of two free parameters in the NFW profile. However, concentration is a descriptive parameter that simply tells us about the shape of the present-day halo profile. In this paper, we wish instead to study the impact of physical properties on halo profiles. We, therefore, study the impact of MAHs on mass profiles directly, without adopting $c_{\rm vir}$ as a proxy.  

Galaxy-galaxy lensing is potentially the most direct method for tracing the mass profiles of dark matter halos in observations. Lensing studies of galaxies or clusters often assume that samples have been selected purely by mass. However, a selection by mass may result in an unintentional selection by MAR. For example, \citet{bradshaw20} showed that richness selected samples might be biased towards young halos with larger recent accretion than other halos of similar mass. As another example, a selection of blue galaxies may result in a selection of halos with higher present MAR compared to average. Until recently, lensing measurements did not have enough signal-to-noise for the impact of secondary halo properties to be of great importance. As lensing surveys continue to expand, it is timely to thoroughly investigate the effect of secondary halo properties on the galaxy-galaxy lensing profile. 

\begin{figure}
    \centering
    \includegraphics[width=\columnwidth]{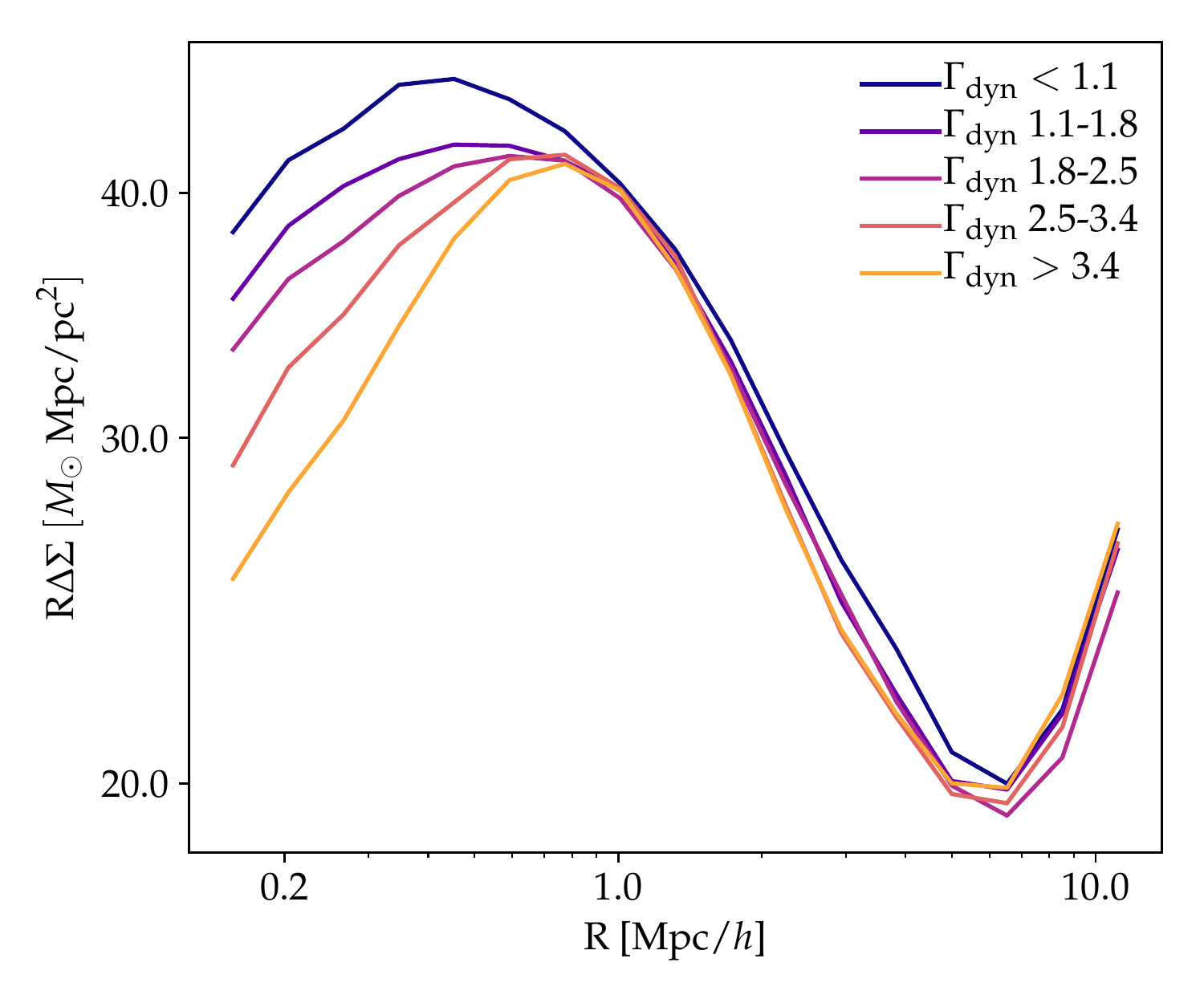}
    \caption{Weak lensing profiles of dark matter halos binned by halo mass accretion rate at fixed mass. The shape of the lensing profile depends on MAR.}

    \label{fig:fig_previous_paper}
\end{figure}

This paper introduces the idea that the whole galaxy-galaxy lensing profile contains information about secondary halo properties, including MAR. Another goal is to show that full weak lensing profiles have better constraining power on MAR than the information contained in the location of halo outskirts \citep[e.g.,][]{xhakaj19, Diemer2014DependenceRate}. Figure 5 of \citet{xhakaj19} (also displayed in Figure \ref{fig:fig_previous_paper}) showed that MAR has a strong impact on the shape of $\ds$. In this paper, we follow up on that analysis by studying how different secondary halo properties affect the shape of the weak lensing profile. Then, we examine whether MAR is detectable with the full area of Hyper Suprime Cam survey  \citep[HSC,][]{Aihara2018FirstProgram}. Given that there is no analytical expression for the variation of the galaxy-galaxy lensing profile with MAR, we build a model based on N-body simulations. We use an emulation technique to predict the weak lensing profiles given halo mass and MAR. We study how constraints are affected by observational scatter in proxies of halo mass and MAR. Then, we analyze how cluster-finding systematics such as miscentering can bias detections of MAR. Finally, we compare our method for detecting MAR with the methods introduced in \citet{Diemer2014DependenceRate} and  \citet{xhakaj19}. We conclude that full shape profile fitting results in better constraints on MAR than the splashback radius alone. 

This paper is structured as follows. We introduce
the numerical simulations and the emulator in Sections \ref{sims} and \ref{methods}. Section \ref{s:secondary_halo_parameters} describes secondary halo parameters that correlate with the shape of the weak lensing profile. We show results in Section \ref{results} and discuss them in Section \ref{discussion}. Finally, we summarize and conclude in Section \ref{summary}. We adopt the cosmology of the MultiDark Planck 2 simulation \citep[MDPL2,][]{Prada2011HaloCosmology}, namely, a flat, $\Lambda \mathrm{CDM}$ cosmology with  $\Omega_\mathrm{\rm m} = 0.307$, $\Omega_\mathrm{\rm b} = 0.0482$, $\sigma_8 = 0.829$, $h = 0.678$, $n_\mathrm{\rm s}=0.9611$, corresponding to the best-fit Planck cosmology \citep{Ade2014PlanckParameters}.

\section{Numerical Simulations and Cluster Sample}\label{sims}

\subsection{Numerical Simulation}\label{numerical_sims}

We study cluster-sized halos from the publicly available MDPL2 simulation, which follows $3840^3$ particles in a cubic box of size $1 \mathrm{Gpc}/h$. Halos are identified using the {\rockstar} halo finder \citep{Behroozi2013TheCores}. Merger trees are constructed using {\fontfamily{lmtt}\selectfont Consistent Trees}, an algorithm that models the evolution of halos among different snapshots \citep{Behroozi2013GravitationallyCosmology}. We select host halos at $z=0.36$ and $\mass \in [10^{13.5}, 10^{14.5}]$ $h^{-1}{\rm M}_\odot$ where $\mass$ is the halo mass enclosed in an overdensity of 200 times the mean matter density of the Universe. We choose this mass range because of the availability of high-signal-to-noise weak lensing measurements around samples of  Luminous Red Galaxies \citep[LRGs, e.g.,][]{lowlensing-alexie, singh19,lange21} and clusters \citep[e.g.,][]{Chang2017TheProfiles}. In Section \ref{full_profile_modelling} we build an emulator to model weak lensing profiles over this mass range. 

In addition, we remove a small number of halos with negative $\mar$ values, where $\mar$ is MAR defined in Equation \ref{eq:mar_mar_definition_sp}. A negative value for MAR can be caused by mergers or fly-bys. After these cuts, halos have $\mar$ values ranging from 0 to 8. Our final set consists of 89790 halos. To compute lensing signals, we use a downsampled catalog of 50 million particles (see Section \ref{lensing}).

\subsection{Fiducial Mock Cluster Sample}\label{fiducial_sample}

We now consider a fiducial sample of clusters drawn from a more narrow mass and MAR range than in the previous section. This baseline cluster sample consists of halos with $\mass \in [10^{13.8}, 10^{14.1}]$ $h^{-1}{\rm M}_\odot$ and $\mar \in [2.5, 3.5]$ at  $z=0.36$ (a total of 4932 halos). We use this sample to predict signals for a fiducial HSC-like cluster sample. The sample that we consider would be drawn from the HSC final survey area, covering 1000 deg$^2$, and $0.3<z<0.6$. This HSC-like sample would contain 1776 clusters. 

We generate lensing error bars for the HSC cluster sample using the same methodology as \citet{Singh2017Galaxy-galaxyProperties}. These forecast error bars include all the terms needed to describe a Gaussian covariance and assume perfect observational tracers for mass and MAR. However, they do not account for effects due to selection or survey masks or the non-gaussian covariance. This results in an overestimated signal to noise by $\sim$25\%, so our predictions will be somewhat optimistic in this regard. For a redshift range of $0.3<z<0.6$, the final HSC volume is 0.36 Gpc$^3/h^3$, which is smaller than the volume of MDPL2 (1 Gpc$^3/h^3$). Because of this, we neglect sample variance in MDPL2 when generating models.

We build two fiducial mock cluster samples: in the first sample, clusters are perfectly centered. The other set consists of clusters that have been off-centered by some distance $\rmis$, where $\rmis$ is defined in Section \ref{model_miscentering}. Section \ref{model_miscentering} also describes how we forward-modeled miscentering in simulated halos. In the following sections, we subdivide this fiducial sample into smaller bins. HSC weak lensing errors for each sub-division are recomputed with the appropriate number of halos in each bin.

\section{Methods}\label{methods}

\subsection{Measuring $\Delta\Sigma$ for the Fiducial Cluster Sample}\label{lensing}

Galaxy-galaxy lensing is defined as the excess surface mass density ($\ds$) and is computed as follows:
\begin{equation}
    \Delta\Sigma(R) = \frac{1}{\pi R^2} \int^R_0 2\pi r \Sigma(r) \mathrm{d}r - \Sigma(R),  
\end{equation}
where $\Sigma(R)$ is the surface mass density projected along the line of sight. In MDPL2, we compute galaxy-galaxy lensing profiles through the excess surface density of dark matter particles in cylinders surrounding host halos. This is implemented in the {\fontfamily{lmtt}\selectfont mean\_delta\_sigma} function of {\halotools} \citep{Hearin2017ForwardHalotools, mean_ds}. 

\subsection{The Dependence of the Shape of $\Delta\Sigma$ on Secondary Halo Properties}\label{cross_correlation}

To study the dependence of the shape of weak lensing profile on secondary halo properties, we employ a Gaussian-driven cross-correlation statistic \citep{Chue2018SomeHalos}. Assuming Gaussian distributed properties with a zero mean, one can compute the expected value of $X$ given $Y$ via: 
\begin{equation}
 \langle X | Y \rangle = \langle XY \rangle \langle YY \rangle ^{-1} Y. 
    \label{eq:cross_correlation_general}
\end{equation}
Here, $X$ corresponds to the $\ds$ profiles, while $Y$ corresponds to the secondary halo parameters for each halo (hereafter $\theta$). For the purpose of this paper, we focus on the change of the lensing profile when secondary halo parameters are varied at fixed halo mass. From Equation \ref{eq:cross_correlation_general}, this can be computed as: 
\begin{equation}
    \frac{\partial \langle \Delta\Sigma_\mathrm{norm} | \theta_\mathrm{norm} \rangle}{\partial \theta_\mathrm{norm}} = \langle \Delta\Sigma_\mathrm{norm} \theta_\mathrm{norm} \rangle \langle \theta_\mathrm{norm} \theta_\mathrm{norm} \rangle ^{-1},
    \label{eq:cross_correlation_DS}
\end{equation}
where $\ds$ and $\theta$ for each halo are normalized over the mean of the sample: 
\begin{align}
\label{eqn:ds_theta_norm}
\begin{split}
& \Delta \Sigma_\mathrm{norm} =  \frac{\Delta\Sigma_\mathrm{halo}-\langle \Delta\Sigma \rangle}{\langle \Delta\Sigma \rangle}
\\
& \theta_\mathrm{norm} = \frac{\theta_\mathrm{halo}-\langle \theta \rangle}{\sigma_\theta}.
\\
\end{split}
\end{align}

If Equation \ref{eq:cross_correlation_DS} is computed such that $\theta$ represents an array of {\it all} secondary halo properties, then $\frac{\partial \langle \Delta\Sigma | \theta \rangle}{\partial \theta}$ describes the change in the weak lensing profile over the change in a secondary halo property, while the other halo properties are fixed. However, all secondary halo parameters considered here are tightly correlated with each other. Our goal is to analyze the impact of each parameter on the shape of the galaxy-galaxy lensing profile. Thus, we compute the full derivative, $\frac{\mathrm{d} \langle \Delta\Sigma | \theta \rangle}{\mathrm{d} \theta}$, instead. We set $\theta$ equal to a single halo property rather than an array of properties as in Equation \ref{eq:cross_correlation_DS}. $\frac{\mathrm{d} \langle \Delta\Sigma | \theta \rangle}{\mathrm{d} \theta}$ describes the change in the shape of the weak lensing profile with respect to the change in halo parameters while allowing the rest of the parameters to be free. To avoid capturing correlations of mass and MAR with the shape of the weak lensing profiles, we select halos from a narrow range around $\mass$ $= 10^{14} h^{-1} \mathrm{M}_\odot$. The results for this analysis are shown in Section \ref{res:dependence}.

\subsection{Modeling Weak Lensing Profiles}\label{model_lensing}

We aim to build a model that predicts stacked $\ds$ profiles given halo bins in $\mass$, $\mar$, and halo miscentering. It has long been known that the shape of the lensing signal can be modeled analytically as an NFW profile with varying concentration \citep{Bartelmann_96, wright_brainerd_00}. However, no analytical model captures the dependence of the weak lensing profiles on other secondary halo properties, such as MAR. Thus, we introduce a model for $\ds$ profiles built directly from N-body simulations. Our model takes the form of an emulator trained on MDPL2 halos. A more advanced version of the emulator also includes the effect of miscentering on the shape of the profile. 

The lensing signal has previously been emulated through N-body simulations \citep[i.e. {\dq}, ][]{Nishimichi21}. {\dq} uses a cosmological suite to emulate halo lensing profiles. The main purpose of this emulator is to model the three-dimensional halo-matter correlation function on various cosmological parameters. Assuming spherical symmetry, it then computes the $\ds$ profile via an analytic projection integral. Our emulator is complementary to this effort: while we do not calibrate cosmology dependence, we relax the spherical symmetry assumption. Moreover, we include a dependence on MAR and miscentering effects. Modeling cosmological parameters and MAR by emulating the $\ds$ signal directly would avoid the spherical symmetry assumption and provide more accurate cosmology results. This will be interesting to explore in future work. 

In the following sections, we describe the method employed to implement and test the emulator. We perform a likelihood analysis to determine whether $\mar$ (a MAH parameter defined in Section \ref{mah_parameters}) can be detected with HSC. The results for this analysis are shown in Sections \ref{res: detectability} and \ref{res:miscentering}.

\subsubsection{Modelling Miscentering}\label{model_miscentering}

Miscentering is a systematic effect in current cluster finding algorithms \citep[e.g.,][]{redmapper1, redmapper2, redmapper3}. We believe that miscentering might have a similar impact as secondary halo properties on the shape of the lensing profile. In this section, we describe analytic modeling for miscentering. Then, we show how we forward model it in MDPL2 halos. 

The miscentered weak lensing profile, $\Delta\Sigma$, can be computed as: 
\begin{equation}
    \Delta \Sigma = (1-f_\mathrm{mis}) \Delta\Sigma_0 + f_\mathrm{mis} \Delta\Sigma_\mathrm{mis},
    \label{eq:misc_stacked}
\end{equation}
\noindent
where $\Delta\Sigma_0$ is the profile in the absence of miscentering, $\Delta\Sigma_\mathrm{mis}$ is the weak lensing profile of miscentered clusters, and $f_\mathrm{mis}$ is the fraction of clusters that are miscentered \citep{des_miscentering}. Given a miscentering distance, $R_\mathrm{mis}$, $\Delta\Sigma_\mathrm{mis}$ is: 
\begin{equation}
    \Delta \Sigma_\mathrm{mis}(R|R_\mathrm{mis}) = \int_0^{2\pi}\frac{d\theta}{2\pi}\Delta\Sigma_{0}\Big(\sqrt{R^2 + R_\mathrm{mis}^2 + 2R R_\mathrm{mis~} \mathrm{cos} \theta} \Big).
    \label{eq:ds_mis_given_rmis}
\end{equation}
The profile average across a distribution of $R_\mathrm{mis}$ is then: 
\begin{equation}
   \Delta \Sigma_\mathrm{mis}(R) = \int d R_\mathrm{mis}P(R_\mathrm{mis})\Delta \Sigma_\mathrm{mis}(R|R_\mathrm{mis}),
   \label{eq:ds_mis_all}
\end{equation}
where $P(R_\mathrm{mis})$ is the probability that a cluster is miscentered by $R_\mathrm{mis}$. 

For the purpose of this project, we forward model miscentering on simulated halos. We displace the center of the halo by a given $R_\mathrm{mis}$, where $R_\mathrm{mis}$ ranges between 0 and 1 Mpc$/h$. The angle, $\phi$, at which we apply the displacement is determined by drawing from a uniform distribution between 0 and $2\pi$. We repeat the draw of $\phi$ 1000 times for each miscentering calculation. Given the original coordinates of the halo center, ($x_0, y_0$), the displaced centers are: 
\begin{align}
\label{eqn:xmis_ymis}
\begin{split}
x_\mathrm{mis} &= x_{0~} + R_\mathrm{mis~} \mathrm{cos} \phi,
\\
y_\mathrm{mis} &= y_{0~} + R_\mathrm{mis~}\mathrm{sin} \phi.
\end{split}
\end{align}
\noindent
Knowing $x_\mathrm{mis}$ and $y_\mathrm{mis}$, we can then compute the weak lensing profile of the miscentered halo, $\Delta\Sigma_\mathrm{mis}$. We assign each halo a probability that it is miscentered, $f_\mathrm{mis}$. Finally, we stack $\ds$ profiles of halos within the same $\fmis$ and $\rmis$ range. This corresponds to the final weak lensing profile of the halo, $\Delta \Sigma$, as in Equation \ref{eq:misc_stacked}.

\subsubsection{Full Profile Modelling}\label{full_profile_modelling}

We model the full shape of stacked $\ds$ profiles using Gaussian Processes (GP) trained on MDPL2 halos through the publicly available code {\fontfamily{lmtt}\selectfont GPy} \citep[][]{gpy2014}. A GP is a selection of random variables that have a Gaussian distribution. GPs can be used as a tool to perform probabilistic regression given a training data set. To train a GP, one needs to constrain the covariance matrix of the training set through a kernel matrix. Here we choose the kernel Matern 3/2 ($k_{3/2}$), which is commonly used to smooth across different points in the training set \citep[][]{matern32}. One can compute $k_{3/2}$ as: 
\begin{equation}
    k_{3/2} = (1 + \frac{\sqrt{3} r}{l})~\mathrm{exp} (-\frac{\sqrt{3} r}{l}), 
    \label{gp_kernel}
\end{equation}
where $l$ is a hyperparameter that defines the characteristic length scale, and $r$ is equal to the absolute difference between two training points.  To train a GP, we need to optimize the hyperparameter $l$ for a given training sample.

We train GPs to build two different emulators: one whose only free parameters are $\mar$ and $\mass$ (hereafter the 2D model) and one that includes the miscentering parameters as well (the 4D model). Our training and test sets for both emulators consist of stacked $\ds$ profiles built from 600 bins in $\mar$ and $\mass$ containing 800 halos each. In our sample, 80\% belong in the training set, while the rest serve as the test sample. The training and testing parameter range is $\mass \in [10^{13.5}, 10^{14.5}] \mathrm{M}_\odot/h$, $\mar \in [0, 8]$, $\fmis \in[0.7]$, and $\rmis \in [0, 1] \mathrm{Mpc}/h$. The mean parameter values of each bin are selected through a latin-hypercube sampling of halos in the $\mass$-$\mar$-$\fmis$-$\rmis$ parameter space. We compute stacked $\ds$ profiles through {\halotools} as described in Section \ref{lensing}.

Given that each training bin has an unequal scatter in $\mass$ and $\mar$, we compute the sample variance for each bin through jackknife resampling. While $\mass$ and $\mar$ are intrinsic halo parameters, we forward model the effect of $\fmis$ and $\rmis$ on $\ds$ as described in the previous section. Training a GP requires choosing the best fit hyper-parameter of our training set. We constrain the kernel function with a heteroscedastic regression given the unequal scatter in the training bins. We construct independent GPs for each radial bin of $\ds$. Hence a specific radial bin has its own optimized hyperparameters. This technique ignores the correlation between different radial bins in $\ds$. Although not ideal, this approach is optimal in balancing emulation speed and accuracy.

\begin{figure*}
    \centering
    \includegraphics[width=\textwidth]{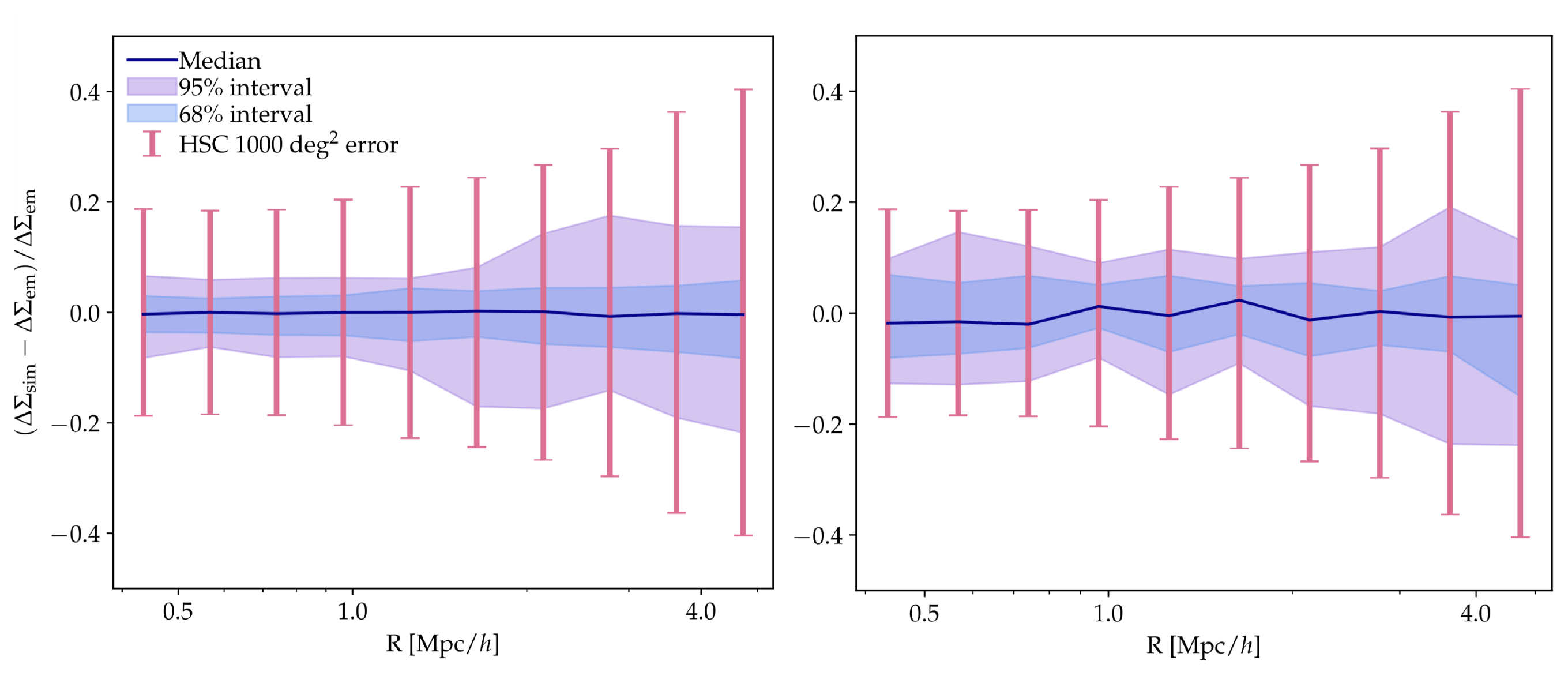}
    \caption{Performance of the 2D (left) and 4D (right) emulators over the respective test samples. Here we measure the deviation of the $\ds$ profile of the test points ($\Delta\Sigma_\mathrm{sim}$) from the respective emulated profiles ($\Delta\Sigma_\mathrm{em}$). Blue and purple contours display the $1\sigma$ and $2\sigma$ errors of the emulator. The vertical error bars display the predicted HSC 1$\sigma$ measurement error for the fiducial cluster sample and the full 1000 deg$^2$ area. The error for both emulators is smaller than the observational error. Specifically, it is 10\% of the measurement error for the 2D emulator and 20\% of the measurement error for the 4D emulator.}

    \label{fig:testing_sample}
\end{figure*}

To test the trained GP, we simulate stacked $\ds$ profiles ($\Delta\Sigma_\mathrm{sim}$) for all bins in the test sample. Then, we study how well the 2D and 4D emulators recover $\Delta\Sigma_\mathrm{sim}$. Figure \ref{fig:testing_sample} displays the error in the 2D and 4D models. For the 2D model, the error in the emulator is 10\% of the measurement error, while for the 4D model, it is 20\% of the measurement error. The modeling error is smaller than the predicted HSC measurement errors for our fiducial cluster sample for both emulators. Thus, the 2D and the 4D emulators are suitable for modeling $\ds$ for this project.

\subsection{Full Profile Fitting}\label{full_profile_fitting}

We follow a Bayesian approach to fit our model to the fiducial stacked lensing profile ($\Delta\Sigma_\mathrm{fid}$). We sample posteriors from the parameter space following a Markov Chain Monte Carlo (MCMC) approach with {\fontfamily{lmtt}\selectfont emcee } \citep{Foreman-Mackey2012Emcee:Hammer}. The likelihood is defined based on a simple $\chi^2$ analysis: 
\begin{equation}
    \mathcal{L}({\bf \delta}) = \frac{-\chi^2({\bf \delta})}{2}.
    \label{eq:likelihood_chi2}
\end{equation}

\noindent Here $\chi^2$ is defined as: 
\begin{equation}
    \chi^2 ({\bf \delta}) = \sum \frac{(\Delta\Sigma_\mathrm{mod}({\bf \delta}) - \Delta\Sigma_\mathrm{fid})^2}{\sigma^2},
    \label{eq:chi2}
\end{equation}
\noindent
where $\Delta\Sigma_\mathrm{mod}$ is evaluated over a set of parameters, $\delta$, using the emulator. First, we model the $\ds$ profile with only the two physical parameters ($\delta$ = [$\mass$, $\mar$]). Then, we add two additional parameters to model miscentering ($\delta$ = [$\mass$, $\mar$, $\fmis$, $\rmis$]). Here $\sigma$ is the sum in quadrature of the anticipated error from the HSC full survey area ($\sigma_{\rm fid}$) and the error of the emulator ($\sigma_{\rm mod}$):
\begin{equation}
 \sigma = \sqrt{\sigma_{\rm fid}^2+\sigma_{\rm mod}^2}.
 \label{eq:total_variance_mcmc}
\end{equation}

Following previous work, we impose Gaussian priors on $\fmis$ and $\rmis$ \citep{rykoff16, Chang2017TheProfiles}. We set the mean of the distribution to our fiducial values for $\fmis$ and $\rmis$, and the width to the respective priors ($\sigma_{\fmis} = 0.11$ and $\sigma_{\rmis} = 0.1 \mathrm{Mpc}/h$ as in \citealt{rykoff16}). Finally, we set flat priors on $\mar$ and $\mass$.  We assess convergence by minimizing the autocorrelation time of the MCMC.

\section{Secondary Halo Parameters}\label{s:secondary_halo_parameters}

We classify secondary halo properties into two sets: instantaneous tracers and MAH parameters. The first set corresponds to parameters that describe the present state of the halo. The second set contains parameters that describe the past MAH of the halo measured on different timescales. 

\subsection{Instantaneous Halo Parameters}\label{instantaneous_parameters}

Instantaneous halo parameters include the ratio of the kinetic to the potential energy of the halo (T/U), the offset between the halo center and its center of mass ($x_{\rm off}$), and halo concentration ($c_\mathrm{vir}$). These parameters can be used as proxies to determine the dynamical state of the halo. For example, halos with low T/U have more potential than kinetic energy than halos with high T/U. This implies that halos with lower T/U are more relaxed. A large value for $x_{\rm off}$ implies that the center of mass is misaligned with the center of the halo. This scenario is possible when the halo experiences (or recently experienced) a major merger and is likely to indicate that the system is in a perturbed state. Finally, the concentration of a halo is defined as $c_\mathrm{vir}=R_\mathrm{vir}/R_\mathrm{s}$ where $R_\mathrm{vir}$ is the virial radius and $R_\mathrm{s}$ is the scale radius. It has been shown that halo concentration correlates with the past MAH \citep[][]{wechsler02, ludlow13}. In particular, for massive halos, larger values of $c_\mathrm{vir}$ at fixed halo mass, on average, correlate with earlier formation times. Thus, we can think of $c_\mathrm{vir}$ as an instantaneous halo parameter that correlates with the past MAH.

\subsection{Mass Accretion History Parameters}\label{mah_parameters}

These parameters quantify the MAHs of halos on different timescales. In principle, halos with lower MAR have stopped growing, so they tend to be older and more relaxed. {\fontfamily{lmtt}\selectfont Consistent Trees} measures MAR over different timescales \citep{Behroozi2013GravitationallyCosmology}. The resulting quantities are contained in {\rockstar} catalogs \citep[][]{Behroozi2013TheCores}. Below we introduce {\rockstar} definitions. Hereafter, we denote MAR as $\Gamma$, with a subscript indicating the timescale over which it is measured. Parameters found in {\rockstar} catalogs will be marked with $``*"$.  

\begin{itemize}
    \item  $\Gamma^*_\mathrm{Inst}$: instantaneous MAR
    \item $\Gamma^*_\mathrm{100Myr}$: MAR averaged over the past 100 Myr
    \item $\Gamma^*_\mathrm{1dyn}$: MAR averaged over the past virial dynamical time
    \item $\Gamma^*_\mathrm{2dyn}$: MAR averaged over the past two virial dynamical times
    \item $\Gamma^*_\mathrm{Mpeak}$: Growth rate of ${\rm M_{peak}}$ (peak mass over all MAH), averaged from $z_1$ to $z_1+0.5$ where $z_1$ is the snapshot under consideration.
\end{itemize}

In addition, we also use the MAH parameter in {\sparta}, which is measured over one dynamical time \citep{Diemer2017TheAlgorithm}.

{\sparta} and {\rockstar} compute MARs in different ways. {\rockstar}'s definition of MAR is: 
\begin{equation}
\Gamma^*_\tau= \frac{\Delta M_\mathrm{vir}}{\tau},
\label{eq:mar_definition_rs}
\end{equation}

\noindent where $\Delta M_\mathrm{vir}$ is the change in the virial mass measured over a given timescale, $\tau$ (i.e. 100 Myr, $t_\mathrm{dyn}$, $2t_\mathrm{dyn}$, etc). {\sparta} defines MAR as:
\noindent
\begin{equation}
    \Gamma_\mathrm{dyn} = \frac{\Delta \mathrm{log} M_{\rm 200b}}{\Delta \mathrm{log} a}.
    \label{eq:mar_mar_definition_sp}
\end{equation}
\noindent
Additional differences in the measurements of {\sparta} and {\rockstar} are explained in \citet{Xhakaj_2019}.

\section{Results}\label{results}

We first study how the shape of the $\ds$ profile depends on secondary halo properties beyond halo mass. We then analyze if $\mar$ is detectable given forecasted HSC lensing errors and varying value of scatter for observational tracers of $\mar$. Finally, we highlight a degeneracy between $\mar$ and miscentering parameters.

\subsection{Dependence of $\Delta\Sigma$ on Secondary Halo Properties}\label{res:dependence}

We start by studying the correlation between the shape of the galaxy-galaxy lensing profile with secondary halo properties. In Section \ref{s:secondary_halo_parameters}, we divided these parameters into two groups: instantaneous tracers (present halo properties that correlate with its dynamical state) and MAH properties (parameters that measure MAR on different timescales). 

We consider a cluster sample selected in the narrow mass range of $M_\mathrm{200m} \in [10^{13.8}, 10^{14.1}] h^{-1}\mathrm{M_\odot}$. We bin this halo sample by each of the secondary halo properties discussed in Section \ref{s:secondary_halo_parameters}. To avoid any bias due to halo mass, we apply the binning methodology introduced in \citet{conditional_binning} and implemented in {\halotools}. We weigh the value of each of the secondary halo properties by halo mass. These weights are uniformly distributed between 0 and 1. We rank the weights and bin them into percentiles. This method guarantees that all bins have the same mass distribution.

\begin{figure*}
    \centering
    \includegraphics[width=.9\textwidth]{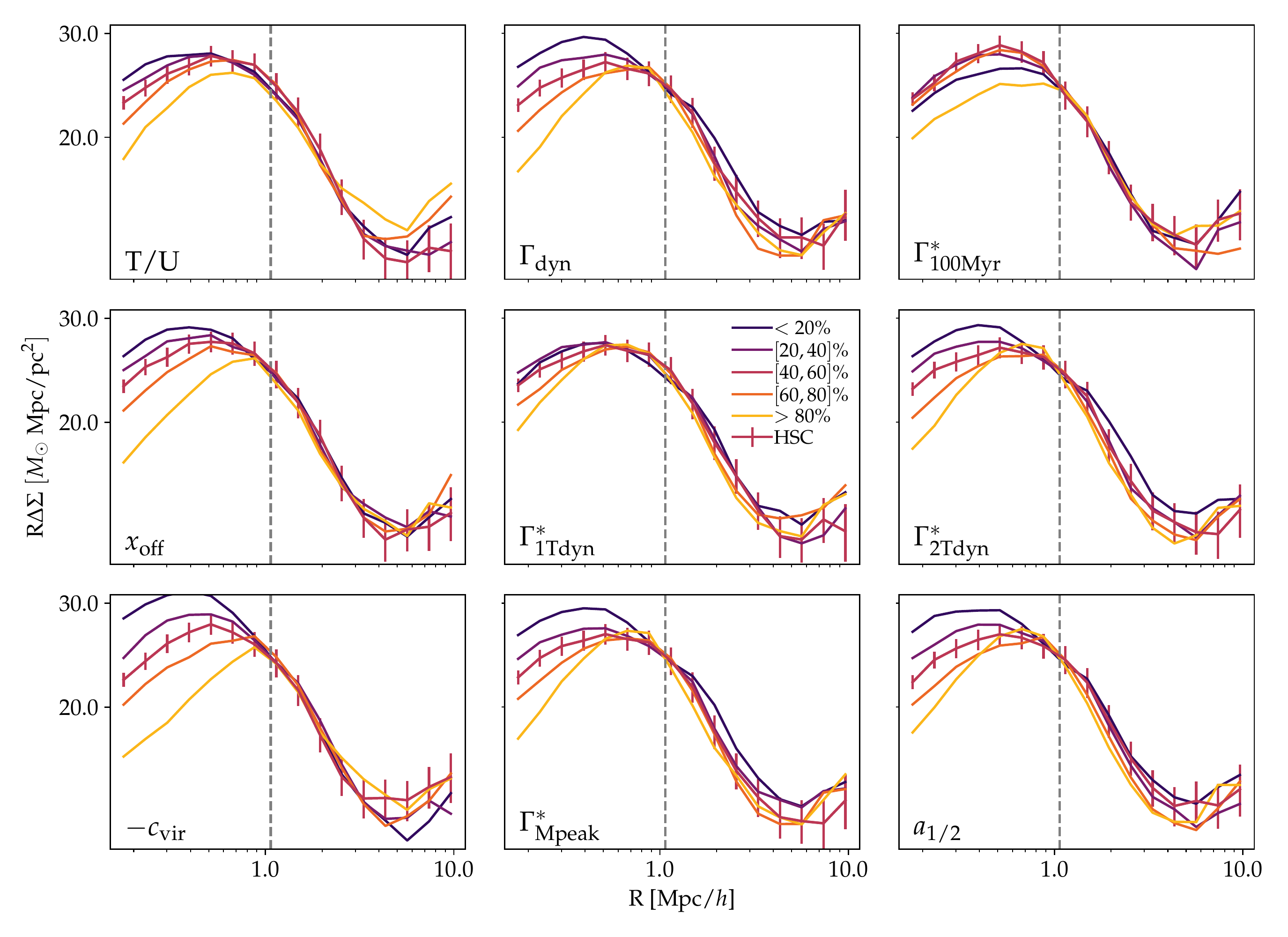}
    \caption{Impact of secondary halo parameters on the shape of the weak lensing profile, $\Delta\Sigma$. Each panel displays variations in $\Delta\Sigma$ with secondary halo properties at fixed halo mass ($\mass$ = $10^{14} M_\odot/h$). Halos are binned by the value of the secondary property using percentiles conditioned on mass as described in Section \ref{res:dependence}. For example, in the upper left, orange corresponds to the 20th percentile of halos with the highest values of T/U, and purple corresponds to the 20th percentile of halos with the lowest values of T/U. The left column shows instantaneous parameters. The middle and right columns display MAH parameters. Anticipated errors from the 1000 deg$^2$ HSC wide-field surveys are overlaid as a reference on each of the panels. The dashed line displays the mean $R_\mathrm{200b}$ of the halo sample. Different halo properties lead to different $\ds$ profiles, and these differences are especially pronounced in the one-halo regime. Upcoming lensing surveys will have the statistical capability to detect secondary halo properties using the shape they imprint on the one-halo lensing signal.}
    \label{fig:ds_profiles_all_pars}
\end{figure*}

Figure \ref{fig:ds_profiles_all_pars} displays the impact of each secondary halo parameter on the shape of the galaxy-galaxy lensing. Secondary halo properties have a substantial impact on $\Delta\Sigma$ in the one-halo regime. All instantaneous parameters show a strong correlation with the inner regions. From the MAH parameters, $\mar$ and $a_{1/2}$ have the most considerable impact. Furthermore, the dependence of the shape on the secondary halo parameters is significant given the predicted HSC lensing errors corresponding to the final survey area. This shows that the impact of secondary halo properties on halo profiles may be detectable with current generation lensing surveys. For this paper, we focus only on the inner profiles ($R <$ 10 Mpc$/h$). Larger simulations would be required to study large-scale assembly bias adequately, and this aspect is left for future work. 

\begin{figure*}
    \centering
    \includegraphics[width=.9\textwidth]{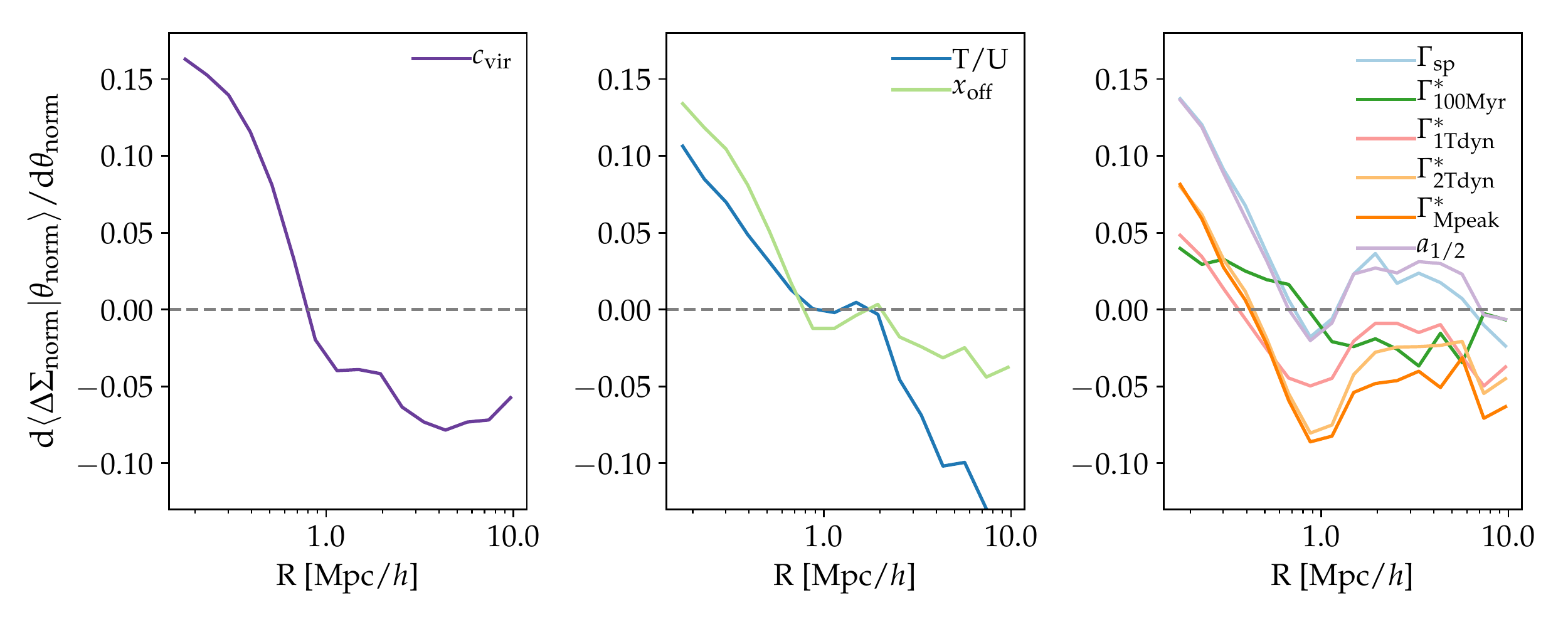}
    \caption{Variance of the normalized $\ds$ profile with respect to a given secondary halo parameter, $\theta$. The $y$-axis can be thought of as the fractional change in $\Delta\Sigma$ for a 1$\sigma$ change in the secondary halo parameter. The left panel displays $c_\mathrm{vir}$. The middle panel displays other instantaneous halo parameters. The right panel displays MAH parameters. All parameters except T/U show stronger variations in the inner regions than in the outer regions. These variations can reach up $16 \%$ in the inner regions. The largest variations are on $R=0.2-0.4 ~\mathrm{Mpc}/h$ scales. Among instantaneous parameters, $c_\mathrm{vir}$ displays the largest variations. Among MAR parameters, $\mar$ and $a_{1/2}$ have the strongest impact.}
    \label{fig:pca_analysis}
\end{figure*}

To quantify the relation between the shape of the $\ds$ profiles and secondary halo parameters, we apply the Gaussian cross-correlation statistic introduced in Section \ref{cross_correlation}. The result of this analysis is displayed in Figure \ref{fig:pca_analysis}. Here we compute the full derivative of $\ds$ over secondary halo parameters. This is the change in the lensing profile when one parameter is changed while allowing the rest to be free, as described in Section \ref{cross_correlation}. Profiles and halo properties are normalized with respect to their mean and standard deviation. The y-axis in Figure \ref{fig:pca_analysis} represents the variance of the normalized $\ds$ profile for a given $1\sigma$ variation of the secondary halo parameter, $\theta$.

Figure \ref{fig:pca_analysis} shows that no single parameter impacts the shape of the lensing profile equally on all scales. Instead, all parameters show a stronger correlation with the inner rather than the outer part of the lensing profile. This correlation is the strongest for $c_\mathrm{vir}$. This is not surprising since, by definition, $c_\mathrm{vir}$ modifies the inner shape of the NFW profile. However, we are more interested in the other parameters, which are physical (in contrast to $c_\mathrm{vir}$ which simply describes the shape of the halo) and are likely to play an important role in determining the properties of galaxies in clusters. Among the instantaneous physical halo parameters, T/U has the largest impact reaching 14\% variations at $R = 0.2~\mathrm{Mpc}/h$. Among MAH physical halo parameters, $\mar$ and $a_{1/2}$ have the largest impact reaching 14\% variations at $R = 0.2~\mathrm{Mpc}/h$.  

\begin{figure*}
    \centering
    \includegraphics[width=.9\textwidth]{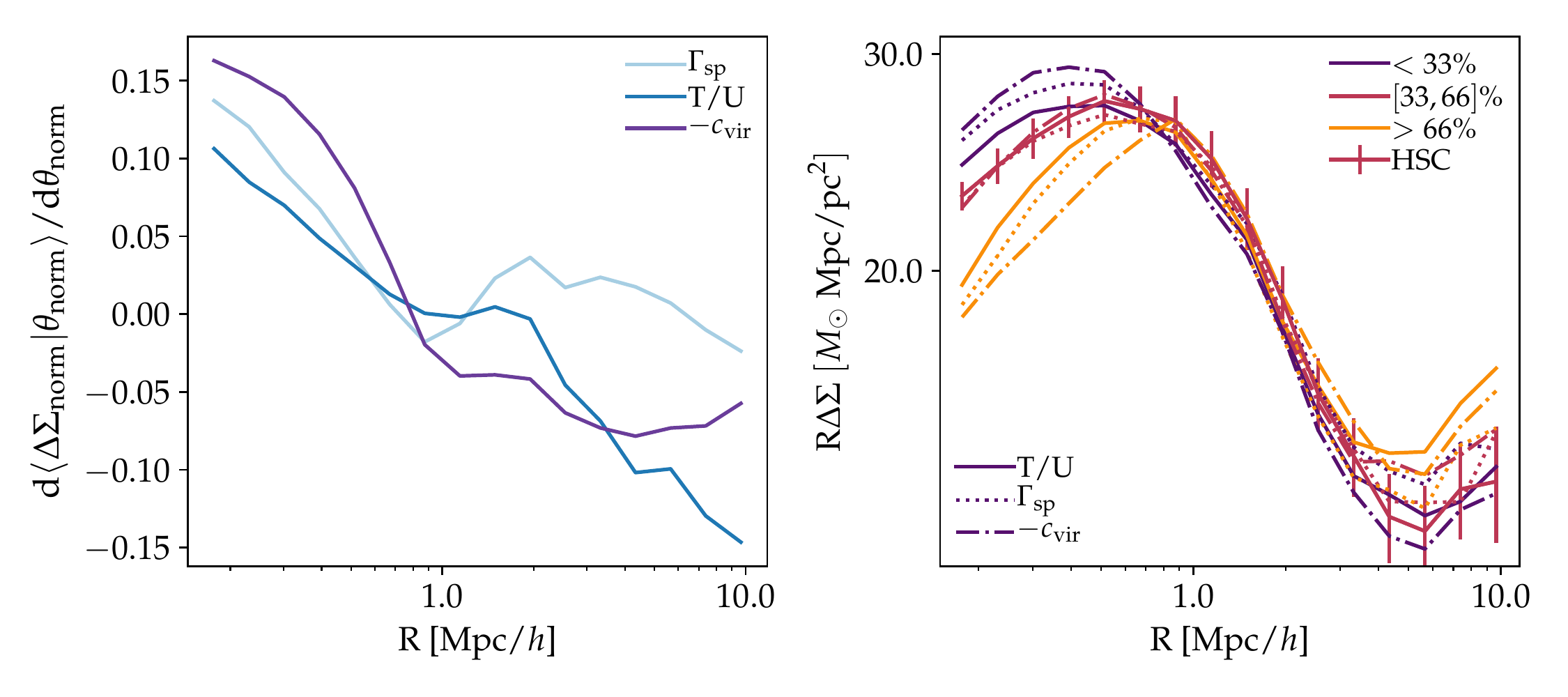}
    \caption{Halo properties that have the strongest impact on the shape of the lensing profile. Left: Variance of the normalized $\ds$ profiles with respect to secondary parameters (Figure \ref{fig:pca_analysis}). Right: Illustration of how the $\ds$ profile changes when we bin by each parameter. Of the three parameters plotted here, $c_\mathrm{vir}$ displays the largest impact in the one-halo regime, followed by $\mar$ and then T/U. This is also illustrated in the right panel, in which the $\ds$ profiles show the largest variance when binned by $c_\mathrm{vir}$. Each parameter also imparts a distinct radial dependence on the shape of $\Delta\Sigma$.}
    \label{fig:top3profs}
\end{figure*}

Figure \ref{fig:top3profs} compares variations in $\Delta\Sigma$ for the parameters that have the largest impact. The left panel displays the variance of the normalized $\ds$ profiles with respect to each secondary parameter. The right panel illustrates the $\ds$ profiles binned by each parameter. Figure \ref{fig:top3profs} reveals two important pieces of information. First, one of the physical parameters with the largest impact on $\Delta\Sigma$ is $\mar$. This is important because hydrodynamic simulations indicate that MAR also correlates with the properties of galaxies in clusters, or the state of the hot gas that is often used to detect clusters, either via X-rays or the tSZ effect \citep[][]{neslon14_nthermal_pressure, shi15_nthermal_pressure}. Second, while each secondary parameter considered here has a roughly similar impact on $\Delta\Sigma$, they have \emph{distinct radial signatures}. This raises the exciting possibility of using $\Delta\Sigma$ to understand which of these secondary halo parameters has the strongest correlations with observed cluster properties (e.g., galaxy content or hot gas).

Having shown that secondary halo properties substantially impact the one-halo $\Delta\Sigma$ profile, we now focus our attention more specifically on one secondary halo property. We choose to focus on $\mar$ because a) $\mar$ is likely to correlate with galaxies and gas in clusters, and b) $\mar$ is one of the parameters that has the most substantial impact in Figure \ref{fig:top3profs}. In the following section, we study whether or not variations in $\ds$ and $\mar$ at fixed halo mass are detectable with HSC. Whether or not these variations can be disentangled from variations due to other secondary halo parameters is left for future work.

\subsection{Can $\mar$ be Measured with Gravitational Lensing?}\label{res: detectability}

In this section, we study if $\mar$ can be detected using the shape of the lensing profile while ignoring miscentering. We first consider the ideal case in which perfect proxies (zero scatter) are available for $\mar$ and $\mass$. We then study the more realistic case in which proxies only yield noisy estimates of $\mar$ and $\mass$.

\subsubsection{Ideal Observational Tracers}\label{res:ideal_tracer}

We first examine the detection of MAR when considering a perfect proxy for $\mass$ and $\mar$. Assuming perfectly centered clusters, we proceed with the full profile fitting routine introduced in Section \ref{full_profile_fitting}. We fit stacked $\ds$ profiles of halos within a narrow mass range of $\mass$ = $10^{14} \mathrm{M}_\odot/h$ and different fiducial values of $\mar$ = [0.78, 2.24, 3.73, 5.54]. The last $\mar$ bin has a larger range due to the small number of halos in that parameter space region. Finally, we use predicted HSC error bars corresponding to the final survey area to constrain the model.

\begin{figure*}
    \centering
    \includegraphics[width=\textwidth]{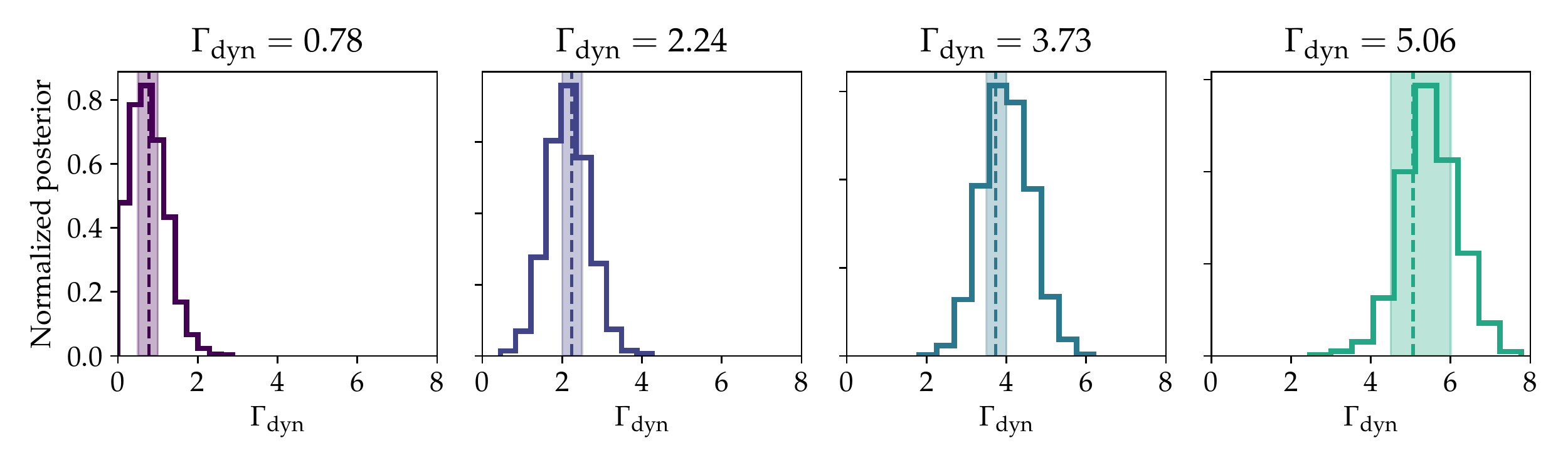}
    \caption{Constraints on $\mar$ using full shape fitting, assuming a 1000 deg$^2$ HSC weak lensing survey, and a perfect (zero scatter) observational tracer of $\mar$. Shaded regions show the range of the fiducial $\mar$ selection. These samples all have the same mass distribution ($\mass$ = $10^{14} \mathrm{M_\odot}/h$), but have different mean values of $\mar$ (dashed lines). Solid lines show the posteriors for $\mar$ using full shape fitting with the emulator. This technique accurately recovers the value of $\mar$ for all four samples.}
    \label{fig:ideal_tracers}
\end{figure*}

Figure \ref{fig:ideal_tracers} shows the posteriors of $\mar$ for the different fiducial bins. The posterior distributions of $\mar$ have a mean and 1$\sigma$ spread of $\Gamma_\mathrm{dyn} = [0.73^{+ 0.49}_{-0.41},~2.16^{+ 0.52}_{-0.51},~4.00^{+ 0.61}_{-0.59},~5.46^{+ 0.72}_{-0.61}]$. The emulator recovers the fiducial values of $\mar$ within 1$\sigma$. Thus, assuming ideal observational tracers for the two halo properties, our full profile fitting technique successfully detects MAR.

\subsubsection{Imperfect Observational Tracers of Halo Mass and $\mar$}\label{res:scatter}

To detect the impact of MAR on cluster profiles, we would need an observational tracer for MAR. However, no observational tracer is a perfect proxy for either halo mass or MAR. Instead, any tracer will be a noisy estimate of halo mass or MAR, which must be considered. This section aims to study the impact of this scatter on the $\ds$ profile. We focus here on detecting variations in MAR at fixed halo mass (which would be a more difficult correlation to detect than the measurement of the average value of MAR at fixed halo mass). We will explore how high this observational scatter can be before it renders profile differences undetectable. 

We first consider the scatter in halo mass. We adopt the mass-richness relation from Table 4 in \citet[][]{McClintock18}:

\begin{equation}
    \langle M \textbar \lambda, z \rangle = M_0 \left( \frac{\lambda}{\lambda_0} \right) ^{F_\lambda} \left( \frac{1+z}{1+z_0}\right)^{G_z},
    \label{eq:mass_richness_relation}
\end{equation}
where $\lambda$ is the cluster richness and ($M_0$, $F_\lambda$, $G_z$) are model parameters. The best fit values for these parameters in \citet[][]{McClintock18} are: ${\rm log_{10}}(M_0) = 14.489 \pm 0.011$, $F_\lambda = 1.356 \pm 0.051$ and $G_z = −0.30 \pm 0.30$. Here, $\lambda_0$ and $z_0$ are pivot values of the model that correspond to the median of the richness and redshift of the cluster sample. For our fiducial cluster sample, $\lambda_0 = 31.6$ and $z_0 = 0.36$. Using MDPL2, we generate a sample of clusters in our target mass range with $\lambda$ values as follows. We first populate the mock with $\lambda$ values using the inverse mass to richness relation given in \citet[][]{McClintock18}.  We then select clusters with $25<\lambda<40$. This sample has a mean halo mass of $\langle \mathrm{log }M_\mathrm{200m} \textbar \lambda \rangle= 13.97~\mathrm{M_\odot}/h$ and the halo mass distribution of this sample has a width of $0.21~ \mathrm{M_\odot}/h$.

Next, we consider the scatter in $\mar$ at fixed proxy (denoted $\scatter$). Observational tracers for $\mar$ have not yet been developed, and possible scatter values are unknown. For this reason, we instead consider a range of possible values for the scatter in this tracer while ignoring the covariance between richness and the potential $\mar$ tracer. To study the impact of scatter on the shape of $\Delta\Sigma$, we compare $\ds$ at fixed $\lambda$ for different samples selected by the proxy for MAR while varying $\scatter$.

\begin{figure*}
    \centering
    \includegraphics[width=\textwidth]{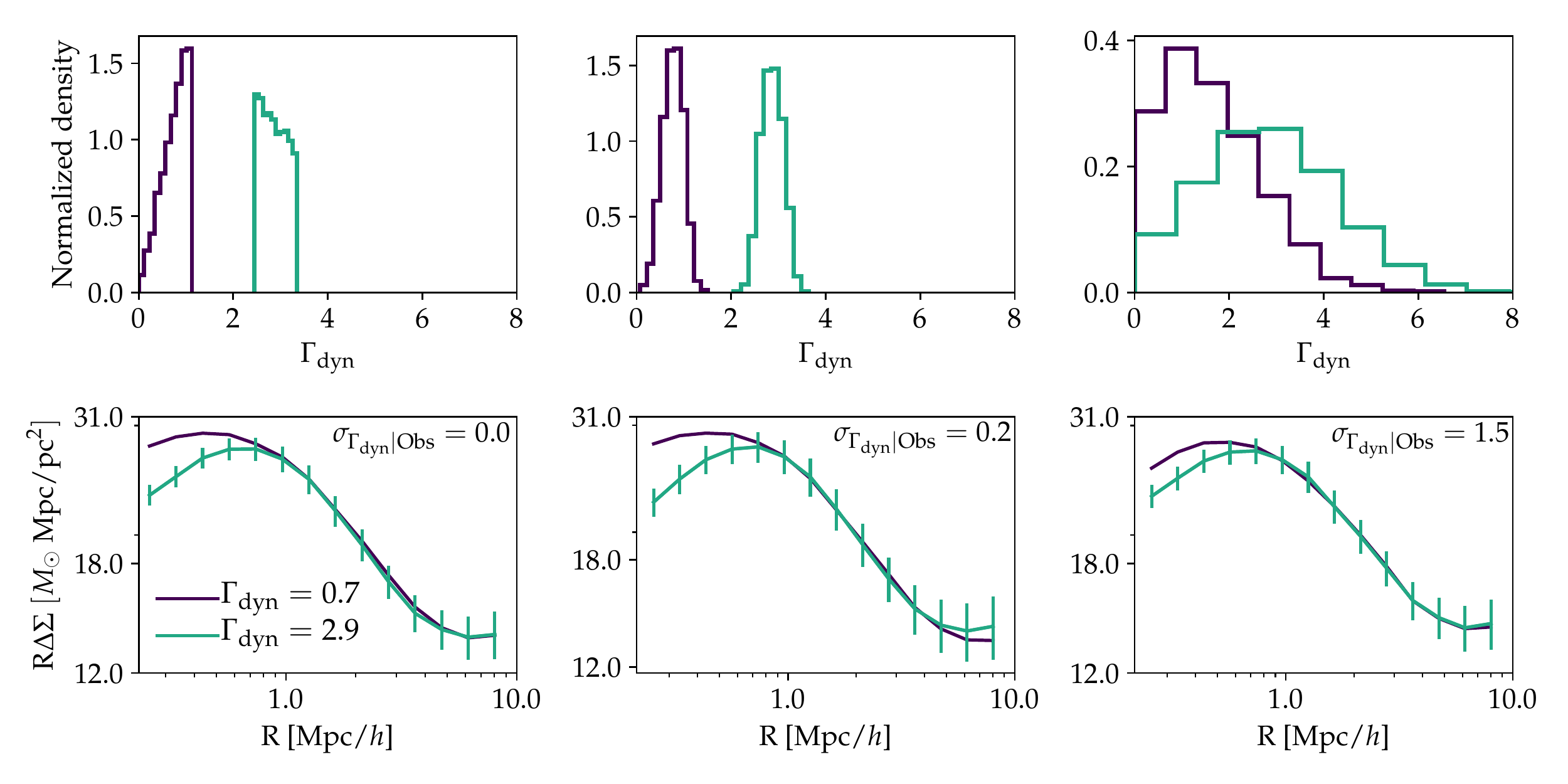}
    \caption{Impact of $\scatter$ on the shape of the $\ds$ profile. Top: true distribution of $\mar$ when the 20th and 80th percentiles are selected using a proxy with zero scatter (left), $\scatter$ = 0.2 (middle) and $\scatter$ = 4. (right). Bottom panels: weak lensing profiles corresponding to these selections. Error bars show the forecasted covariance from the full HSC survey scaled by the number density of the halos in each selection. The $\ds$ signal becomes difficult to distinguish for values of $\scatter$ > 1.5.}
    \label{fig:scatter_effect}
\end{figure*}

Figure \ref{fig:scatter_effect} displays the result of this exercise. As the scatter increases, the inner regions of the lensing profiles become more similar. For low values of $\scatter$, we can differentiate between the shape of the profiles of lower and higher $\mar$ bins given predicted HSC error bars. In contrast, the profiles are indistinguishable for high values of $\scatter$. Therefore, the scatter in $\mar$ has a large impact on the shape of the profile and will be an important factor in determining whether or not variations in $\mar$ can be detected in $\Delta\Sigma$ at fixed halo mass (richness).

\begin{figure*}
    \centering
    \includegraphics[width=0.8\textwidth]{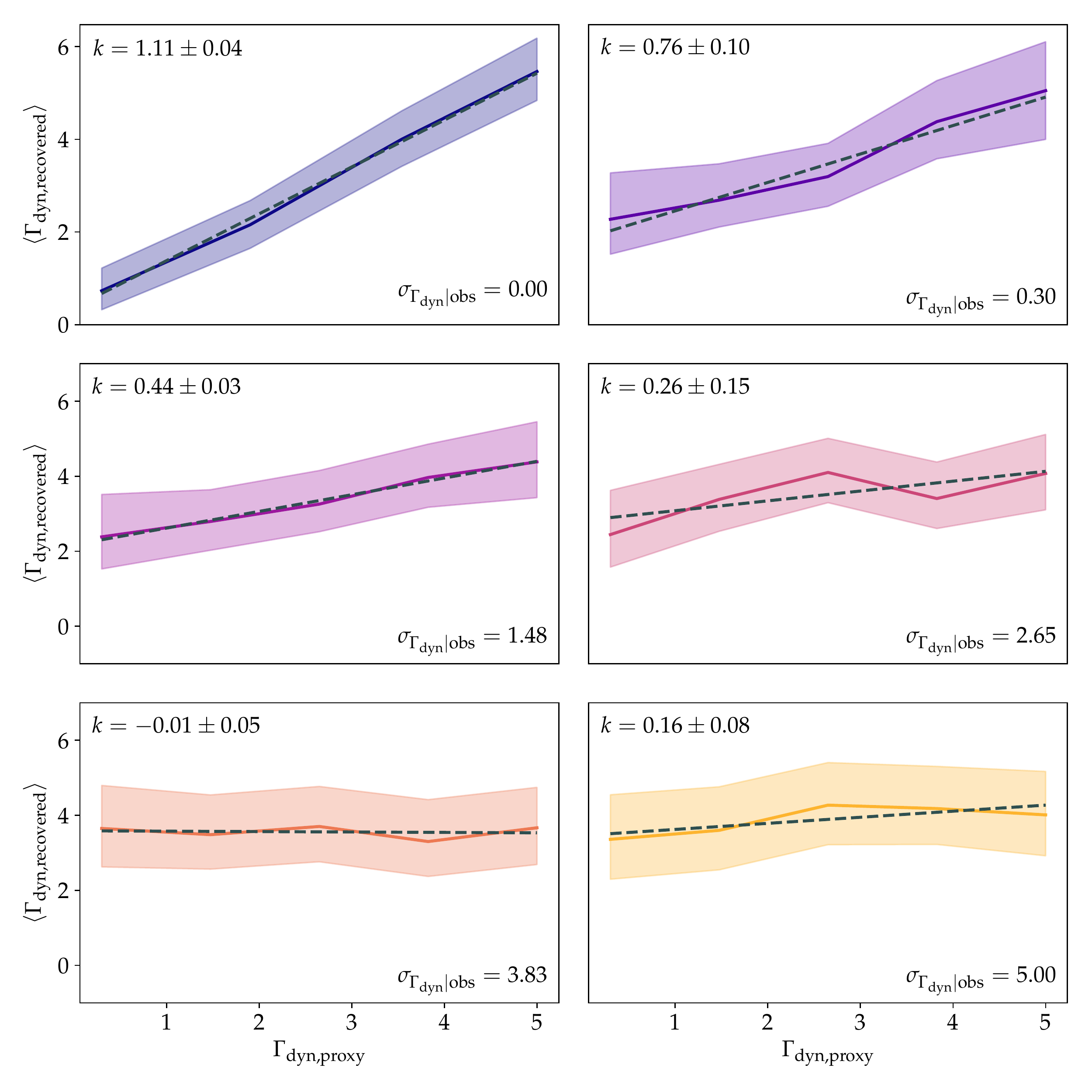}
    \caption{Recovered $\mar$ with respect to proxy $\mar$ for different values of observational scatter ($\scatter$). 
    Best fit slopes ($k$) are displayed on the top right corner of each panel. $k=1$ corresponds to the scenario in which we recover the proxy $\mar$. This occurs at $\scatter=0$ (top right panel). For high scatter (bottom panels), $k=0$. At high scatter, the profiles become indistinguishable (also shown in Figure \ref{fig:scatter_effect}) and the trend cannot be detected. Assuming the full HSC survey area, we can expect to detect variations in lensing profiles due to $\mar$ at fixed halo mass at $3\sigma$ significance if $\scatter < 1.5$.}
    \label{fig:scatter_chi2}
\end{figure*}

Next, we compute the maximum value of $\scatter$ for which we can detect a trend in  $\mar$ at fixed halo mass when fitting the lensing profiles with the 2D emulator. We start by selecting halos with $\lambda \in [25, 40]$, where $\lambda$ is assigned to each halo as described above. Then, we randomly draw halos from Gaussian distributions with a given mean $\mar$ and scatter. We vary the mean $\mar$ from 0.3 to 7 and $\scatter$ from 0 to 5. Then, we build $\ds$ profiles for all samples. Predicted HSC lensing error bars are scaled based on the expected number of halos in each bin. We fit the respective lensing profile with our 2D emulator, whose free parameters are halo mass and MAR. Figure \ref{fig:scatter_chi2} displays the recovered trends for different values of $\scatter$. 

At zero scatter the correlation between the recovered and proxy $\mar$ is linear with a slope of 1. Thus at zero scatter, the emulator recovers the proxy $\mar$ (also displayed in Figure \ref{fig:ideal_tracers}). As the scatter increases, the linear correlation between the recovered and proxy $\mar$ flattens. This is because lensing profiles have become more similar with increasing scatter (also shown in Figure \ref{fig:scatter_effect}). Finally, at high scatter (bottom panel), the slope of the correlation is zero. Regardless of the proxy $\mar$, the recovered $\mar$ is constant across all bins. This agrees with Figure \ref{fig:scatter_effect}, which shows that for high scatter, the $\ds$ profiles for high and low MAR are indistinguishable. 

To quantify the maximum $\scatter$ at which we can hope to detect $\mar$, we perform a linear fit of correlation at each value of scatter. We then measure the maximum $\scatter$ for which the slope is greater than 0 at $3\sigma$ significance. This corresponds to $\scatter = 1.5$. We conclude that a proxy for MAR with $\scatter < 1.5$ is required to detect MAR in real data. This analysis accounts for the scatter between halo mass and richness but ignores possible correlations between tracers for MAR and $\lambda$. Possible avenues for constructing a tracer for  $\mar$ are discussed in Section \ref{disc:obs_proxies}.

\subsection{Full Profile Fitting versus Halo Outskirts}\label{res:fullprofile_vs_outskirts}

 \citet{Diemer2014DependenceRate} proposed a new analytical model (hereafter DK14) to fit median and mean halo density profiles with better than 10\% accuracy. The DK14 model expresses halo density profiles as comprised of two parts: a truncated Einasto profile, describing the collapsed region, and a power-law term, describing the infalling region of the halo. The steepening of the Einasto profile occurs on the outskirts of the halo. Its specific location is called the truncation radius ($r_\mathrm{t}$) and is a free parameter in the model. \citet{Diemer2014DependenceRate} showed that there was a tight correlation between $r_\mathrm{t}$ and MAR, which was modeled as an exponential: 
 
\begin{equation}
r_\mathrm{t}/R_\mathrm{200b} = 0.62+1.18e^{-\mar/1.5},
\label{eq:rt_gamma_dk14}
\end{equation}
where $R_\mathrm{200b}$ is the halo radius (assuming an overdensity of 200 times the mean matter density of the Universe). Due to different simulations used in DK14 and this work, we choose to recalibrate Equation 17 with MDPL2. The new  $r_\mathrm{t} - \Gamma_\mathrm{dyn}$ relation for our simulation is: 
\begin{equation}
r_\mathrm{t}/R_\mathrm{200m} = 0.71 + 2.16e^{-\Gamma_\mathrm{dyn}/0.5}. 
\label{eq:rt_gamma_ourwork}
\end{equation}

This section compares constraints on MAR when using the scaling relation from DK14 with those obtained from full profile fitting. We use a sample of mock clusters with perfect centering and with $\mass$ $\in [10^{13.8},10^{14.1}] \mathrm{M}_\odot/h$ and $\mar$ $\in [2.5, 3.5]$. We constrain the free parameters of each model, assuming HSC full survey errors. We adopt a Bayesian approach to fit the lensing profile with both DK14 and our model. Finally, we sample the posterior of $r_\mathrm{t}$ with an MCMC analysis implemented in {\fontfamily{lmtt}\selectfont emcee}. Given that DK14 does not measure $\mar$ directly, we compute the posterior of $\mar$ with Equation \ref{eq:rt_gamma_ourwork}. 

\begin{figure}
    \centering
    \includegraphics[width=\columnwidth]{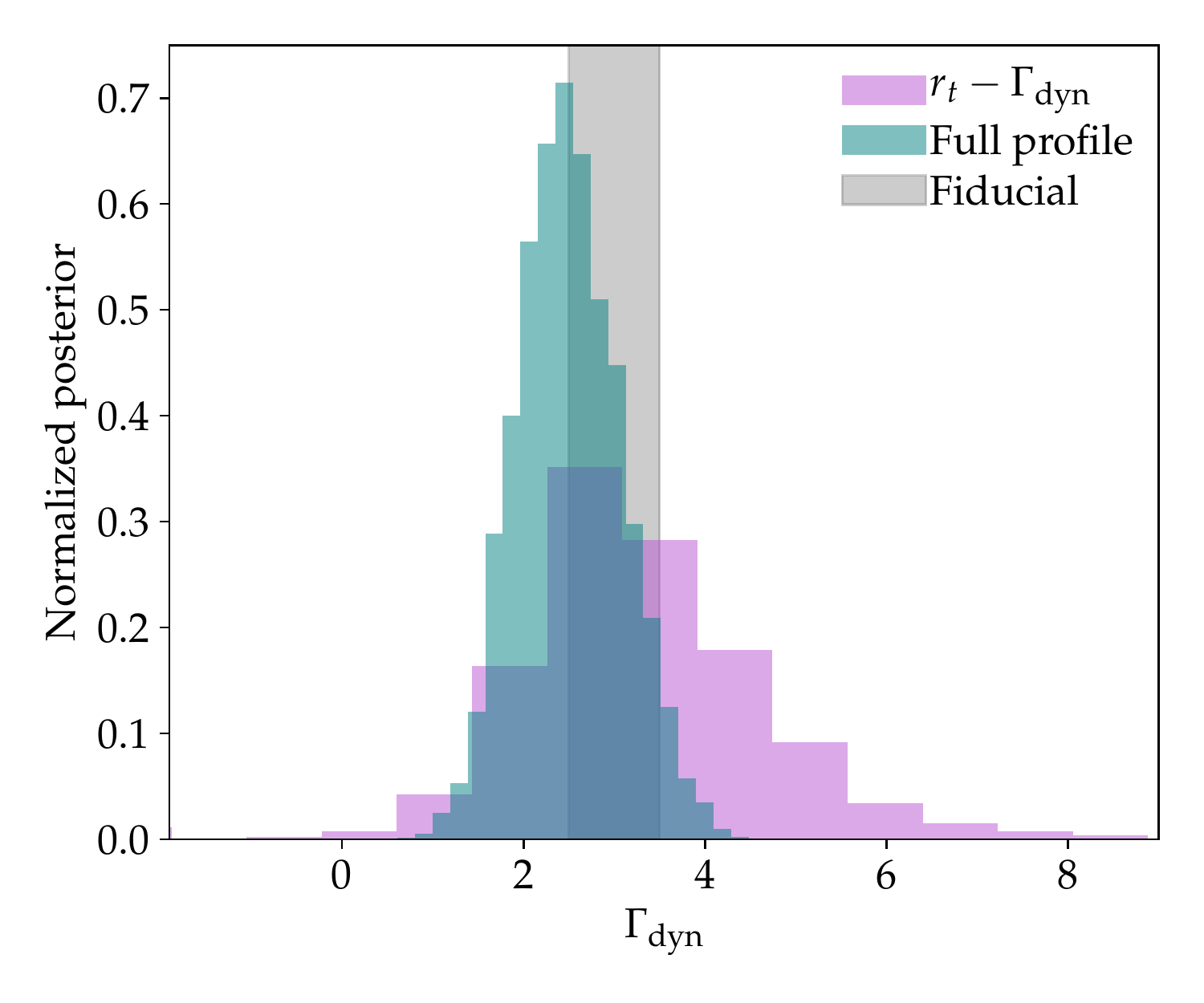}
    \caption{Comparison of constraints on $\mar$ using full shape fitting versus the location of halo outskirts via the DK14 $r_\mathrm{t} - \Gamma_\mathrm{dyn}$ scaling relation. They grey shaded region shows the range of $\mar$ values in the sample. Full shape fitting yields an error on $\mar$ of $\sigma_{\Gamma_\mathrm{dyn}}=1.1$ whereas the location of halo outskirts yields $\sigma_{\Gamma_\mathrm{dyn}}=2.4$. This is because the full $\Delta\Sigma$ profile contains more information about $\mar$ than the $r_\mathrm{t} - \Gamma_\mathrm{dyn}$ scaling relation.}
    \label{fig:dk14_emulator}
\end{figure}

Figure \ref{fig:dk14_emulator} compares MAR constraints from the $r_\mathrm{t}$-$\mar$ scaling relation and full profile modeling technique. It is clear that our method gives tighter constraints on $\mar$ than the $r_\mathrm{t} - \Gamma_\mathrm{dyn}$ relation. The $r_\mathrm{t}$-$\mar$ scaling relation constrains MAR with an error of $\sigma_{\Gamma_\mathrm{dyn}} = 2.4$. In contrast, full profile fitting yields $\sigma_{\Gamma_\mathrm{dyn}} = 1.1$. The full profile fitting method is 2.1 times more constraining than the $r_\mathrm{t} - \Gamma_\mathrm{dyn}$ relation. The difference between the two methods relates to the length of their fiducial data vectors. The $r_\mathrm{t} - \Gamma_\mathrm{dyn}$ relation only uses information from a narrow radial range whereas full shape fitting uses all radial scales of $\ds$. This is why this method yields tighter constraints on $\mar$.

\subsection{Impact of Miscentering}\label{res:miscentering}

\begin{figure}
    \centering
    \includegraphics[width=0.9\columnwidth]{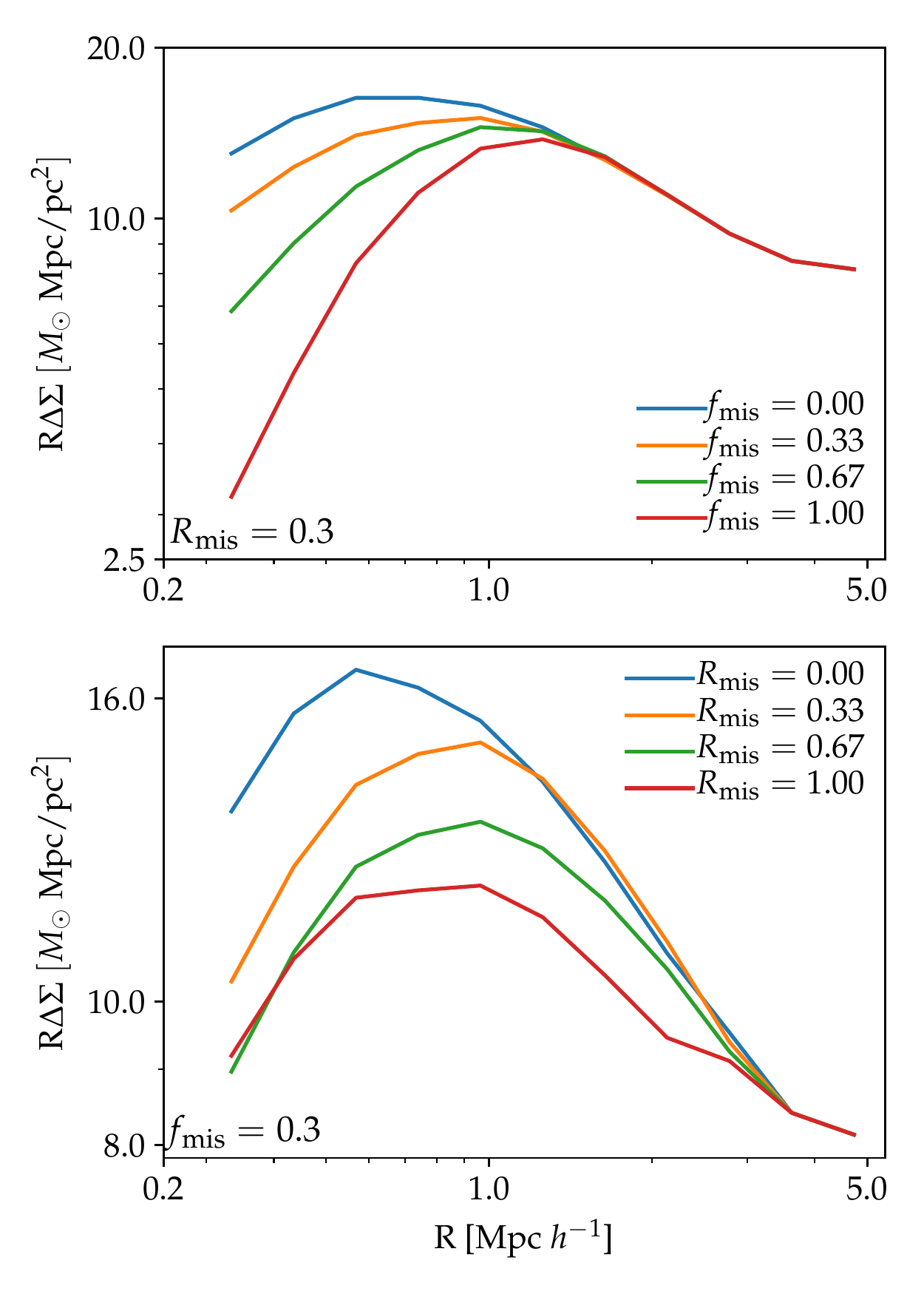}
    \caption{Effect of miscentering on the shape of the $\ds$ profile while varying $\fmis$ (top) and $\rmis$ (bottom). Both parameters impact the one halo regime in a similar fashion as $\mar$ (compare with Figure \ref{fig:ds_profiles_all_pars}). Therefore, miscentering and  MAR are likely to be correlated.}
    \label{fig:wl_mis}
\end{figure}

Miscentering is a systematic effect introduced when the central galaxy of the cluster is misidentified. This translates into an artificial change in the shape of the $\ds$ profile. One can model the effect of cluster miscentering through two free parameters, $\fmis$ and $\rmis$, as described in Section \ref{model_miscentering}. Figure \ref{fig:wl_mis} displays the impact of these two parameters on the shape of $\ds$ for halos with mean $M_{\rm{200m}} = 10^{14}  ~h^{-1}\mathrm{M}_\odot$ and mean $\Gamma_\mathrm{dyn} = 0.67$. The top panel shows how the profile changes as we increase the fraction of miscentered halos at fixed $R_\mathrm{mis} = 0.3~h^{-1} \mathrm{Mpc}$. The bottom panel shows the impact of the miscentering distance at fixed $f_\mathrm{mis}= 0.3$. Both panels indicate that miscentering introduces changes in the inner part of the $\ds$ profile. It is important to notice that $\fmis$ causes variations in $\Delta\Sigma$ similar to those caused by variations in $\mar$ (see Figure \ref{fig:ds_profiles_all_pars}). While the impact of $\rmis$ is qualitatively different from $\mar$, low values of $\rmis$ still affect the shape of the profile on smaller scales. Thus, we expect $\fmis$ and $\rmis$ to be degenerate at some level with $\mar$. 

\begin{figure*}
    \centering
    \includegraphics[width=.8\textwidth]{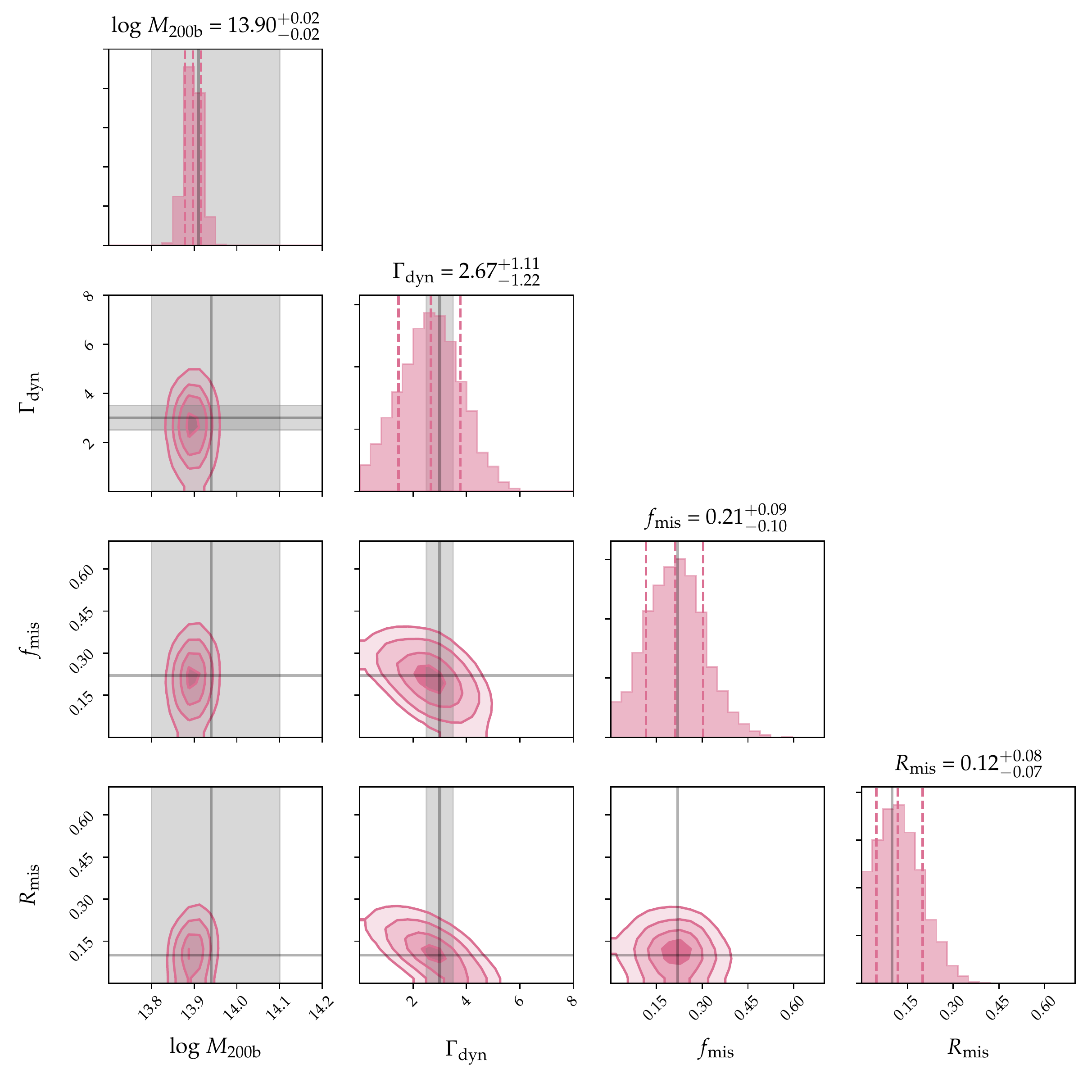}
    \caption{Posteriors of full profile fitting for a set of clusters with mean $\mar=3$ and miscentered assuming $\rmis=0.1 ~\mathrm{Mpc}/h$ and $\fmis= 0.22$. Grey shaded regions indicate the cuts on $\mass$ and $\mar$ that define the fiducial sample ($\mass \in [13.8, 14.1]$ and $\mar \in[2.5, 3.5]$). Solid grey lines correspond to the mean halo values and the true miscentering values.  Dashed lines indicate the one sigma contours of the posterior distribution. Whereas halo mass and MAR are uncorrelated, there is a strong correlation between $\mar$ and the two miscentering parameters. This is because miscentering and $\mar$ have a similar impact on the shape of the one-halo term.}
    \label{fig:corner_4d_mis}
\end{figure*}

We include $\fmis$ and $\rmis$ as two additional parameters in our emulator to quantify this degeneracy. We fit the fiducial data vector of the halo sample that contains miscentered halos. The fiducial parameters are $\mass$ $\in [10^{13.8},10^{14.1}] \mathrm{M}_\odot/h$, $\mar$ $\in [2.5, 3.5]$, $\fmis$=0.22, and $\rmis$=0.1 $h^{-1}{\rm Mpc}$. The values for the two miscentering parameters are justified by the priors used for modeling miscentering in \citet{rykoff16} and \citet{Chang2017TheProfiles}. Finally, the parameters are constrained through the predicted HSC error bars. Figure \ref{fig:corner_4d_mis} shows the results of the fit. We recover all four fiducial parameters within $1\sigma$. The marginalized posteriors of $\mar$ with $\fmis$ and $\rmis$ in Figure \ref{fig:corner_4d_mis} imply that $\mar$ is correlated with both miscentering parameters as predicted from Figure \ref{fig:wl_mis}. If not modeled properly, this degeneracy can potentially impact our physical understanding of halos. 

The fact that miscentering and $\mar$ parameters are correlated has two implications. First, neglecting the impact of miscentering could bias attempts to measure $\mar$. Second, if cluster samples are selected by halo mass and $\mar$, but the $\mar$ selection is ignored, then miscentering parameters could be biased. We now study these two effects.

\begin{figure}
    \centering
    \includegraphics[width=\columnwidth]{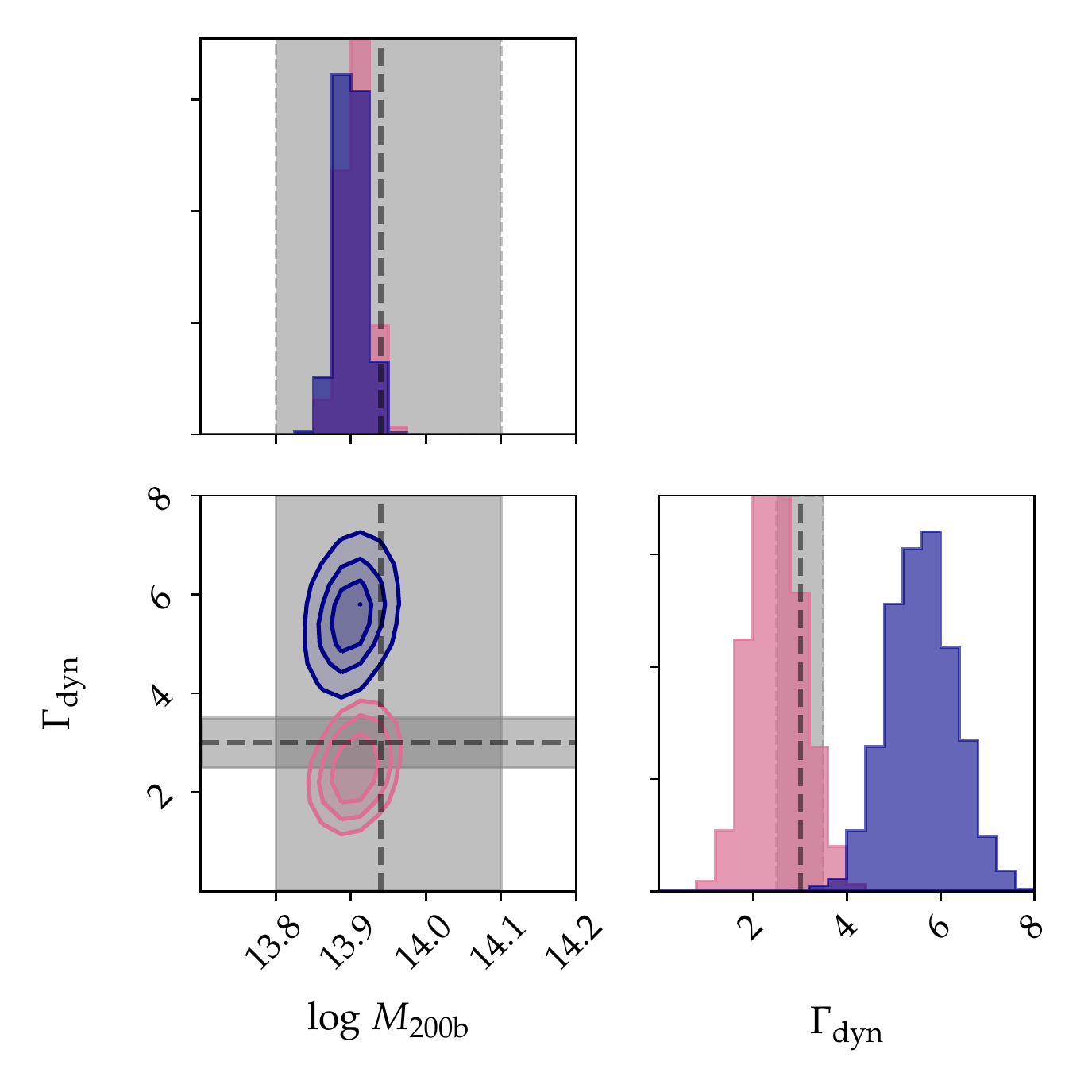}
    \caption{Impact of ignoring miscentering when fitting for $\mar$. Here we assume a fiducial halo sample with $\mass \in [13.8, 14.1]$ and $\mar \in[2.5, 3.5]$. Shaded regions display the  $\mass$ and $\mar$ cuts that define the sample. Dashed lines are the mean values. Pink contours show posteriors when the sample is perfectly centered. In this case, halo properties are recovered accurately. Blue contours show posteriors when clusters are miscentered with $\fmis=0.3$ and $\rmis=0.3$ $\mathrm{Mpc}/h$, but miscentering is ignored when performing the fit. Ignoring miscentering biases the recovery of $\mar$. }
    \label{fig:corner_2d}
\end{figure}

To study the impact of miscentering on the detection of $\mar$, we fit the same fiducial set of clusters as above ($\mass = 10^{14} \mathrm{M}_\odot$ and $\mar \in [2.5, 3.5]$). We apply miscentering to these clusters assuming $\fmis$ = 0.3 and $\rmis =0.3 \mathrm{Mpc}/h$. Finally, we constrain the parameters with error bars corresponding to the final HSC survey area. Here we aim to study the bias in $\mar$ due to miscentering. We fit both fiducial samples of clusters (perfectly centered and miscentered) with the 2D emulator (which does not model miscentering effects). Figure \ref{fig:corner_2d} shows the results of these fits.  Both samples have the same distribution of $\mass$ and $\mar$, so they should, in principle, recover the same range of parameters. However, while the emulator recovers the true $\mass$ and $\mar$ in the absence of miscentering, it fails to do so for miscentered clusters. Thus, not including miscentering in the model introduces a bias in the detection of $\mar$. 

\begin{figure*}
    \centering
    \includegraphics[width=.8\textwidth]{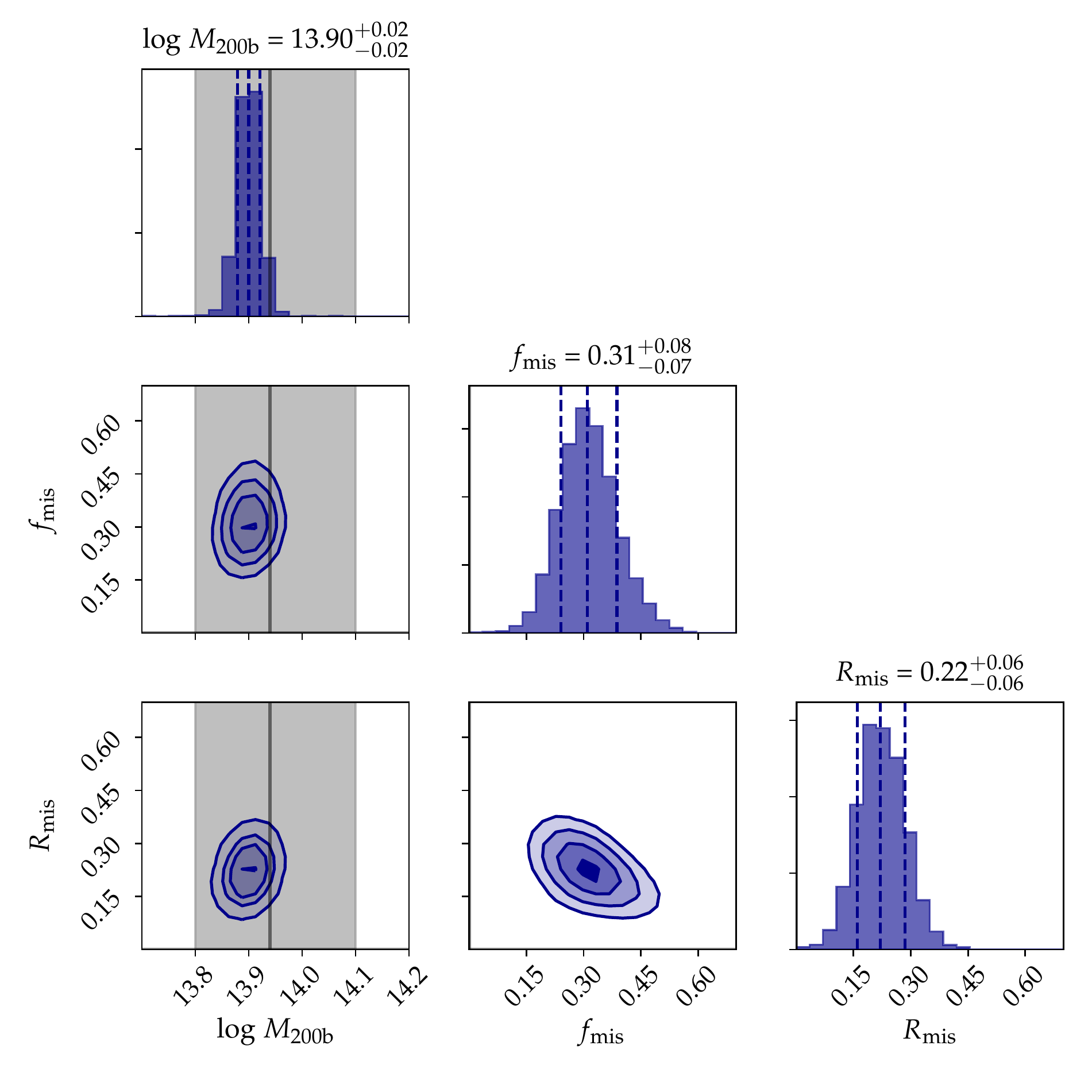}
    \caption{Impact of ignoring $\mar$ when modeling $\ds$ profiles. The fiducial sample contains only perfectly centered clusters such that $\fmis$=0 and $\rmis$ = 0 Mpc/$h$. Grey shaded regions indicate the fiducial $\mass$ range. Dashed lines indicate the one sigma contours of the posterior distribution. We assume perfect tracers of halo mass and MAR. In fitting the fiducial profile, we allow the emulator only to vary $\mass$, $\fmis$, and $\rmis$. Both $\fmis$ and $\rmis$ are overestimated (the true values are $\fmis$=0 and $\rmis$ = 0 Mpc/$h$). This bias comes from not accounting for the effect of MAR on the shape of the profile.}
    \label{fig:corner_3d_mis}
\end{figure*}

Next, we explore the case when the model includes miscentering but not $\mar$. This time we allow for $\mass$, $\fmis$, and $\rmis$ to be the only free parameters in the full profile fitting routine. We fit for the same fiducial samples as above and consider only the sample with perfectly centered clusters. We constrain the parameters with the forecasted error bars of the HSC final survey area. Figure \ref{fig:corner_3d_mis} displays the posteriors of the fit. The emulator overestimates the impact of miscentering in the shape of the profile.
This can be an issue as miscentering is routinely used in modeling $\ds$ profiles, while MAR is not. For a $\mar$ biased selected sample, one could erroneously overrate the impact of miscentering on the profile. Given the results from Figures \ref{fig:corner_2d} and \ref{fig:corner_3d_mis}, it is clear that joint modeling of halo MAR and observational miscentering may be necessary in order to avoid introducing biases when studying weak lensing profiles around clusters.

\section{Discussion}\label{discussion}

\subsection{On the Possibility of Detecting $\mar$ in Real Data}\label{disc:real_data}

\subsubsection{Observational Proxies for $\mar$}\label{disc:obs_proxies}

To constrain MAR in real data, one first needs a low scatter tracer of MAR. Whereas tracers for halo mass have been well studied, tracers for MAR have yet to be developed. Multiple approaches could be used to identify a suitable tracer. One approach would be to study scaling relations from empirical models such as the {\fontfamily{lmtt}\selectfont Universe Machine} \citep{theUM_behroozi},  {\fontfamily{lmtt}\selectfont Emerge} \citep{emerge} etc. The advantage of this approach is that several empirical models are publicly available, and looking for a low scatter tracer of $\mar$ would be a relatively straightforward undertaking. Another approach would be to study the covariance among different halo and cluster properties through hydrodynamic simulations. For example, \citet{farahi2020} use the TNG300 simulation \citep[][]{tng_illustris} to improve the halo mass to cluster property relation. They find that conditioning on the magnitude gap lowers the scatter in the tracer to halo mass relation. One could, in principle, repeat the same analysis for MAR. Finally, a third approach would be to employ Machine Learning (ML) techniques. ML has been proposed as a new tool to connect galaxy and halo properties. One can use unsupervised learning to detect patterns in the data set without assuming a physical model. For example, \citet{galaxynet} introduce {\fontfamily{lmtt}\selectfont GalaxyNet}, a neural network that studies correlations between galaxy and halo properties. By utilizing reinforcement learning, {\fontfamily{lmtt}\selectfont GalaxyNet} is instead trained directly on observed data. The algorithm is rewarded or penalized at specific points during learning, based on how well it recovers the observed statistics. As such, it is not informed about the underlying physics. \citet{galaxynet} demonstrate that ML provides reliable and unbiased methods for finding proxies for halo properties in observations. We will explore such techniques in future work.

\subsubsection{Impact of Baryons}\label{disc:baryons}

Figure \ref{fig:pca_analysis} showed that the shape of $\Delta\Sigma$ varies when secondary halo properties are varied at fixed halo mass. We also showed that the one-halo regime could also be impacted by cluster finding systematics such as miscentering (Figure \ref{fig:wl_mis}). This leads to a degeneracy between MAR and miscentering (Section \ref{res:miscentering}). Baryonic effects can influence the shape of the profile at scales below a few Mpc. \citet{lowlensing-alexie} and \citet{lange21} show that baryons impact the one-halo regime ($R<4$ $\mathrm{Mpc}/h$) and can modify $\Delta\Sigma$ by up to 20\% \citep[see Figure 12 in][]{lowlensing-alexie}.

While the effect of baryons on the shape of the lensing signal has been well studied, the interplay between baryonic effects and MAR is not well known. In future work, it would be interesting to examine if the effect of baryons on the shape of the profile is degenerate with the impact of MAR or if the two effects have distinct radial signatures that allow them to be disentangled. Given that our full profile modeling technique is trained on dark matter only simulations, it does not consider baryonic effects. In the present work, we cannot test whether there is a degeneracy between MAR and baryonic feedback.

\subsection{Subhalo Cross-Correlations}\label{disc:subhalos}

While galaxy-galaxy lensing is an excellent tracer of the total mass distribution in halos, a joint analysis with subhalo cross-correlations could help improve constraints on MAR. The use of subhalo cross-correlations has become more popular for constraining halo properties \citep[e.g.,][]{More2016DetectionClusters, Chang2017TheProfiles, Baxter2017}. While including subhalo cross-correlations is likely to tighten constraints on $\mar$,  modeling subhalos adds complexity to the analysis for two reasons. First, subhalos in simulations are prone to numerical artifacts such as artificial disruption \citep{numericalartifacts}. Therefore, a robust way to identify them in simulations and ensuring that the relationship between subhalo profiles and $\mar$ at fixed halo mass is preserved will be critical. Second, the relationship between the galaxy selection function and the subhalo profiles needs to be adequately modeled. For example, \citet{Baxter2017} showed that the distributions of red and blue galaxies within a cluster were different. Red galaxies were concentrated inside the cluster, implying that these galaxies had been quenched after one or more orbits in the cluster. Blue star-forming galaxies, on the other hand, were concentrated in the infalling region of the cluster. Finally, other effects such as tidal stripping of subhalos and mass segregation also need to be considered. Future work will tackle these caveats and explore the possibility of using galaxy cross-correlations as observational probes to detect MAR.
    
\section{Summary and Conclusions}\label{summary}

In this paper, we use the MDPL2 simulation to study the impact of secondary halo properties on the shape of the $\Delta\Sigma$ lensing profile at fixed halo mass. We also study whether or not it will be possible to detect the relationship between halo mass and $\mar$ using the full HSC weak lensing data set. To achieve these goals, we build an emulator trained on MDPL2 that models $\ds$ given halo mass, MAR, and miscentering parameters. Our main findings are: 

\begin{itemize}
    \item All secondary halo properties considered in this paper impact the shape of $\Delta\Sigma$ on scales below 10 Mpc$/h$. The impact of these parameters is significant compared to the predicted errors on $\Delta\Sigma$ assuming a 1000 deg$^2$ HSC survey and clusters drawn from $0.3<z<0.6$ (Figure \ref{fig:ds_profiles_all_pars}). While each secondary parameter has a roughly similar impact on $\Delta\Sigma$, they have \emph{distinct radial signatures} (Figures \ref{fig:pca_analysis} and \ref{fig:top3profs}). This raises the exciting possibility of using $\Delta\Sigma$ to understand which of these secondary halo parameters has the strongest correlations with cluster properties such as galaxy content or hot gas.
    
    \item Among those parameters considered, $c_\mathrm{vir}$ has the largest impact on $\Delta\Sigma$. However, this is by definition since $c_\mathrm{vir}$ characterizes variations in the shape of the NFW profile. Instead, in this paper, we focused more specifically on the impact of physical halo parameters on $\Delta\Sigma$. This is important because galaxies or hot gas in clusters are likely to correlate with physical secondary halo properties (such as halo accretion rate over one dynamical time, $\mar$). Among physical parameters, T/U, $\mar$, and $a_{1/2}$ have the largest impact on the shape of $\ds$ with variations reaching up to 14\%  at $R=0.2 \mathrm{Mpc}/h$ (Figure \ref{fig:pca_analysis}).

    \item We then investigate whether or not trends in $\mar$ at fixed halo mass could be detected with the full 1000 deg$^2$ HSC survey. Assuming that a perfect observational tracer for $\mar$ is available, we show that a cluster sample drawn from $0.3<z<0.6$ could be used to successfully detect $\mar$ (Figure \ref{fig:ideal_tracers}).
    
    \item We further consider the impact of scatter between an observational tracer and $\mar$. The scatter on $\mar$ impacts the smaller scales of $\ds$ (Figure \ref{fig:scatter_effect}). We show that a proxy with $\scatter<1.5$ is required to detect $\mar$ (Figure \ref{fig:scatter_chi2}) with the full HSC survey.
    
    \item We compare constraints on $\mar$ obtained from full shape fitting with the emulator to those obtained from the DK14 $r_\mathrm{t}$-$\mar$ scaling relation. We find that the full profile fitting routine constrains $\mar$ 2.1 times better than the $r_\mathrm{t}$-$\mar$ scaling relation (Figure \ref{fig:dk14_emulator}). 
    
    \item Finally, we study the interplay between miscentering and MAR. We find that the effect of miscentering on $\ds$ is degenerate with MAR (Figure \ref{fig:corner_4d_mis}). This means that miscentering needs to be taken into account for an unbiased detection of $\mar$ (Figure \ref{fig:corner_2d}). Similarly, to avoid biased estimates of miscentering parameters, the impact of $\mar$ should be carefully considered (Figure \ref{fig:corner_3d_mis}). This is important because it is likely that common cluster selection techniques (such as richness or X-ray selections) may preferentially select clusters according to $\mar$ in addition to mass. 
 
\end{itemize}

We have shown that present-day lensing data sets have the statistical capability to place constraints on halo accretion rates. This will open new possibilities for studying correlations between galaxy and cluster properties and secondary halo properties such as $\mar$. Observational insights into these correlations may lead to new methods for constraining the growth rate \citep[][]{hurier19} and will help improve semi-analytic and empirical models of galaxy formation. Three main challenges will need to be tackled to carry out this program. First, it will be necessary to identify and characterize a low scatter proxy for $\mar$. Second, it will be essential to understand the interplay between observational tracers, $\mar$, and other secondary halo properties impacting $\Delta\Sigma$ in different ways. Third, it will be necessary also to consider the impact of baryonic effects. With the Roman Space Telescope \citep[][]{roman}, the Vera Rubin Observatory Legacy Survey of Space and Time \citep[LSST,][]{lsst} and \textit{Euclid}  \citep{Laureijs2011EuclidReport} on the horizon, the coming decade will usher in the potential for even stronger constraints on $\ds$ than those considered here. Techniques for measuring and disentangling secondary halo properties are thus becoming paramount, and the methodology introduced here paves the way for a new generation of lensing-based constraints on the mass and assembly history of dark matter halos.

\section*{Acknowledgements}

EX is grateful to Joe DeRose and Christopher Bradshaw for helpful conversations. This work was wrapped up during the Coronavirus pandemic and would not have been possible without the hard work of all the essential workers who did not have the privilege of working from home. EX is also grateful to her cat for offering emotional support and motivation to advance this project. This research was supported in part by the National Science Foundation under Grant No. NSF PHY-1748958. This material is based on work supported by the UD Department of Energy, Office of Science, Office of High Energy Physics under Award Number DE-SC0019301. AL acknowledges support from the David and Lucille Packard Foundation and the Alfred. P Sloan foundation. EX acknowledges the generous support of Mr. and Mrs. Levy via the LEVY fellowship.

\bibliographystyle{mnras}
\bibliography{citations}

\begin{thebibliography}{}
\makeatletter
\relax
\def\mn@urlcharsother{\let\do\@makeother \do\$\do\&\do\#\do\^\do\_\do\%\do\~}
\def\mn@doi{\begingroup\mn@urlcharsother \@ifnextchar [ {\mn@doi@}
  {\mn@doi@[]}}
\def\mn@doi@[#1]#2{\def\@tempa{#1}\ifx\@tempa\@empty \href
  {http://dx.doi.org/#2} {doi:#2}\else \href {http://dx.doi.org/#2} {#1}\fi
  \endgroup}
\def\mn@eprint#1#2{\mn@eprint@#1:#2::\@nil}
\def\mn@eprint@arXiv#1{\href {http://arxiv.org/abs/#1} {{\tt arXiv:#1}}}
\def\mn@eprint@dblp#1{\href {http://dblp.uni-trier.de/rec/bibtex/#1.xml}
  {dblp:#1}}
\def\mn@eprint@#1:#2:#3:#4\@nil{\def\@tempa {#1}\def\@tempb {#2}\def\@tempc
  {#3}\ifx \@tempc \@empty \let \@tempc \@tempb \let \@tempb \@tempa \fi \ifx
  \@tempb \@empty \def\@tempb {arXiv}\fi \@ifundefined
  {mn@eprint@\@tempb}{\@tempb:\@tempc}{\expandafter \expandafter \csname
  mn@eprint@\@tempb\endcsname \expandafter{\@tempc}}}

\bibitem[\protect\citeauthoryear{{Adhikari}, {Dalal}  \&
  {Chamberlain}}{{Adhikari} et~al.}{2014}]{Adhikari2014SplashbackHalos}
{Adhikari} S.,  {Dalal} N.,   {Chamberlain} R.~T.,  2014, \mn@doi [\jcap]
  {10.1088/1475-7516/2014/11/019}, \href
  {https://ui.adsabs.harvard.edu/abs/2014JCAP...11..019A} {2014, 019}

\bibitem[\protect\citeauthoryear{{Aihara} et~al.,}{{Aihara}
  et~al.}{2018}]{Aihara2018FirstProgram}
{Aihara} H.,  et~al., 2018, \mn@doi [\pasj] {10.1093/pasj/psx081}, \href
  {https://ui.adsabs.harvard.edu/abs/2018PASJ...70S...8A} {70, S8}

\bibitem[\protect\citeauthoryear{{Bartelmann}}{{Bartelmann}}{1996}]{Bartelmann_96}
{Bartelmann} M.,  1996, \aap, \href
  {https://ui.adsabs.harvard.edu/abs/1996A&A...313..697B} {313, 697}

\bibitem[\protect\citeauthoryear{{Baxter} et~al.,}{{Baxter}
  et~al.}{2017}]{Baxter2017}
{Baxter} E.,  et~al., 2017, \mn@doi [\apj] {10.3847/1538-4357/aa6ff0}, \href
  {https://ui.adsabs.harvard.edu/abs/2017ApJ...841...18B} {841, 18}

\bibitem[\protect\citeauthoryear{{Behroozi}, {Wechsler}  \& {Wu}}{{Behroozi}
  et~al.}{2013a}]{Behroozi2013TheCores}
{Behroozi} P.~S.,  {Wechsler} R.~H.,   {Wu} H.-Y.,  2013a, \mn@doi [\apj]
  {10.1088/0004-637X/762/2/109}, \href
  {https://ui.adsabs.harvard.edu/abs/2013ApJ...762..109B} {762, 109}

\bibitem[\protect\citeauthoryear{{Behroozi}, {Wechsler}, {Wu}, {Busha},
  {Klypin}  \& {Primack}}{{Behroozi}
  et~al.}{2013b}]{Behroozi2013GravitationallyCosmology}
{Behroozi} P.~S.,  {Wechsler} R.~H.,  {Wu} H.-Y.,  {Busha} M.~T.,  {Klypin}
  A.~A.,   {Primack} J.~R.,  2013b, \mn@doi [\apj]
  {10.1088/0004-637X/763/1/18}, \href
  {https://ui.adsabs.harvard.edu/abs/2013ApJ...763...18B} {763, 18}

\bibitem[\protect\citeauthoryear{{Behroozi}, {Wechsler}, {Hearin}  \&
  {Conroy}}{{Behroozi} et~al.}{2019}]{theUM_behroozi}
{Behroozi} P.,  {Wechsler} R.~H.,  {Hearin} A.~P.,   {Conroy} C.,  2019,
  \mn@doi [\mnras] {10.1093/mnras/stz1182}, \href
  {https://ui.adsabs.harvard.edu/abs/2019MNRAS.488.3143B} {488, 3143}

\bibitem[\protect\citeauthoryear{{Behroozi}, {Hearin}  \& {Moster}}{{Behroozi}
  et~al.}{2021}]{behroozi_scale_dependence}
{Behroozi} P.,  {Hearin} A.,   {Moster} B.~P.,  2021, arXiv e-prints, \href
  {https://ui.adsabs.harvard.edu/abs/2021arXiv210105280B} {p. arXiv:2101.05280}

\bibitem[\protect\citeauthoryear{{Blumenthal}, {Faber}, {Flores}  \&
  {Primack}}{{Blumenthal} et~al.}{1986}]{blumenthal86}
{Blumenthal} G.~R.,  {Faber} S.~M.,  {Flores} R.,   {Primack} J.~R.,  1986,
  \mn@doi [\apj] {10.1086/163867}, \href
  {https://ui.adsabs.harvard.edu/abs/1986ApJ...301...27B} {301, 27}

\bibitem[\protect\citeauthoryear{{Bradshaw}, {Leauthaud}, {Hearin}, {Huang}  \&
  {Behroozi}}{{Bradshaw} et~al.}{2020}]{bradshaw20}
{Bradshaw} C.,  {Leauthaud} A.,  {Hearin} A.,  {Huang} S.,   {Behroozi} P.,
  2020, \mn@doi [\mnras] {10.1093/mnras/staa081}, \href
  {https://ui.adsabs.harvard.edu/abs/2020MNRAS.493..337B} {493, 337}

\bibitem[\protect\citeauthoryear{{Chang} et~al.,}{{Chang}
  et~al.}{2018}]{Chang2017TheProfiles}
{Chang} C.,  et~al., 2018, \mn@doi [\apj] {10.3847/1538-4357/aad5e7}, \href
  {https://ui.adsabs.harvard.edu/abs/2018ApJ...864...83C} {864, 83}

\bibitem[\protect\citeauthoryear{{Chue}, {Dalal}  \& {White}}{{Chue}
  et~al.}{2018}]{Chue2018SomeHalos}
{Chue} C. Y.~R.,  {Dalal} N.,   {White} M.,  2018, \mn@doi [\jcap]
  {10.1088/1475-7516/2018/10/012}, \href
  {https://ui.adsabs.harvard.edu/abs/2018JCAP...10..012C} {2018, 012}

\bibitem[\protect\citeauthoryear{{Diemer}}{{Diemer}}{2017}]{Diemer2017TheAlgorithm}
{Diemer} B.,  2017, \mn@doi [\apjs] {10.3847/1538-4365/aa799c}, \href
  {https://ui.adsabs.harvard.edu/abs/2017ApJS..231....5D} {231, 5}

\bibitem[\protect\citeauthoryear{{Diemer} \& {Kravtsov}}{{Diemer} \&
  {Kravtsov}}{2014}]{Diemer2014DependenceRate}
{Diemer} B.,  {Kravtsov} A.~V.,  2014, \mn@doi [\apj]
  {10.1088/0004-637X/789/1/1}, \href
  {https://ui.adsabs.harvard.edu/abs/2014ApJ...789....1D} {789, 1}

\bibitem[\protect\citeauthoryear{{Diemer}, {Kravtsov}  \& {More}}{{Diemer}
  et~al.}{2013}]{diemer13}
{Diemer} B.,  {Kravtsov} A.~V.,   {More} S.,  2013, \mn@doi [\apj]
  {10.1088/0004-637X/779/2/159}, \href
  {https://ui.adsabs.harvard.edu/abs/2013ApJ...779..159D} {779, 159}

\bibitem[\protect\citeauthoryear{{Dore} et~al.,}{{Dore} et~al.}{2019}]{roman}
{Dore} O.,  et~al., 2019, \baas, \href
  {https://ui.adsabs.harvard.edu/abs/2019BAAS...51c.341D} {51, 341}

\bibitem[\protect\citeauthoryear{{Fall} \& {Efstathiou}}{{Fall} \&
  {Efstathiou}}{1980}]{fall_80}
{Fall} S.~M.,  {Efstathiou} G.,  1980, \mn@doi [\mnras]
  {10.1093/mnras/193.2.189}, \href
  {https://ui.adsabs.harvard.edu/abs/1980MNRAS.193..189F} {193, 189}

\bibitem[\protect\citeauthoryear{{Farahi}, {Ho}  \& {Trac}}{{Farahi}
  et~al.}{2020}]{farahi2020}
{Farahi} A.,  {Ho} M.,   {Trac} H.,  2020, \mn@doi [\mnras]
  {10.1093/mnras/staa291}, \href
  {https://ui.adsabs.harvard.edu/abs/2020MNRAS.493.1361F} {493, 1361}

\bibitem[\protect\citeauthoryear{{Fillmore} \& {Goldreich}}{{Fillmore} \&
  {Goldreich}}{1984}]{Fillmore1984Self-similarUniverse}
{Fillmore} J.~A.,  {Goldreich} P.,  1984, \mn@doi [\apj] {10.1086/162070},
  \href {https://ui.adsabs.harvard.edu/abs/1984ApJ...281....1F} {281, 1}

\bibitem[\protect\citeauthoryear{{Foreman-Mackey}, {Hogg}, {Lang}  \&
  {Goodman}}{{Foreman-Mackey} et~al.}{2013}]{Foreman-Mackey2012Emcee:Hammer}
{Foreman-Mackey} D.,  {Hogg} D.~W.,  {Lang} D.,   {Goodman} J.,  2013, \mn@doi
  [\pasp] {10.1086/670067}, \href
  {https://ui.adsabs.harvard.edu/abs/2013PASP..125..306F} {125, 306}

\bibitem[\protect\citeauthoryear{{GPy}}{{GPy}}{2012}]{gpy2014}
{GPy} since 2012, {GPy}: A Gaussian process framework in python,
  \url{http://github.com/SheffieldML/GPy}

\bibitem[\protect\citeauthoryear{{Gao}, {Springel}  \& {White}}{{Gao}
  et~al.}{2005}]{gao2005}
{Gao} L.,  {Springel} V.,   {White} S. D.~M.,  2005, \mn@doi [\mnras]
  {10.1111/j.1745-3933.2005.00084.x}, \href
  {https://ui.adsabs.harvard.edu/abs/2005MNRAS.363L..66G} {363, L66}

\bibitem[\protect\citeauthoryear{Genton}{Genton}{2002}]{matern32}
Genton M.~G.,  2002, J. Mach. Learn. Res., 2, 299–312

\bibitem[\protect\citeauthoryear{{Gunn} \& {Gott}}{{Gunn} \&
  {Gott}}{1972}]{Gunn1972OnEvolution}
{Gunn} J.~E.,  {Gott} J.~Richard I.,  1972, \mn@doi [\apj] {10.1086/151605},
  \href {https://ui.adsabs.harvard.edu/abs/1972ApJ...176....1G} {176, 1}

\bibitem[\protect\citeauthoryear{{Hearin} et~al.,}{{Hearin}
  et~al.}{2017}]{Hearin2017ForwardHalotools}
{Hearin} A.~P.,  et~al., 2017, \mn@doi [\aj] {10.3847/1538-3881/aa859f}, \href
  {https://ui.adsabs.harvard.edu/abs/2017AJ....154..190H} {154, 190}

\bibitem[\protect\citeauthoryear{{Hurier}}{{Hurier}}{2019}]{hurier19}
{Hurier} G.,  2019, arXiv e-prints, \href
  {https://ui.adsabs.harvard.edu/abs/2019arXiv190406951H} {p. arXiv:1904.06951}

\bibitem[\protect\citeauthoryear{{Lange}, {Yang}, {Guo}, {Luo}  \& {van den
  Bosch}}{{Lange} et~al.}{2019}]{mean_ds}
{Lange} J.~U.,  {Yang} X.,  {Guo} H.,  {Luo} W.,   {van den Bosch} F.~C.,
  2019, \mn@doi [\mnras] {10.1093/mnras/stz2124}, \href
  {https://ui.adsabs.harvard.edu/abs/2019MNRAS.488.5771L} {488, 5771}

\bibitem[\protect\citeauthoryear{{Lange}, {Leauthaud}, {Singh}, {Guo}, {Zhou},
  {Smith}  \& {Cyr-Racine}}{{Lange} et~al.}{2021}]{lange21}
{Lange} J.~U.,  {Leauthaud} A.,  {Singh} S.,  {Guo} H.,  {Zhou} R.,  {Smith}
  T.~L.,   {Cyr-Racine} F.-Y.,  2021, \mn@doi [\mnras] {10.1093/mnras/stab189},
  \href {https://ui.adsabs.harvard.edu/abs/2021MNRAS.502.2074L} {502, 2074}

\bibitem[\protect\citeauthoryear{{Laureijs} et~al.,}{{Laureijs}
  et~al.}{2011}]{Laureijs2011EuclidReport}
{Laureijs} R.,  et~al., 2011, arXiv e-prints, \href
  {https://ui.adsabs.harvard.edu/abs/2011arXiv1110.3193L} {p. arXiv:1110.3193}

\bibitem[\protect\citeauthoryear{{Leauthaud} et~al.,}{{Leauthaud}
  et~al.}{2017}]{lowlensing-alexie}
{Leauthaud} A.,  et~al., 2017, \mn@doi [\mnras] {10.1093/mnras/stx258}, \href
  {https://ui.adsabs.harvard.edu/abs/2017MNRAS.467.3024L} {467, 3024}

\bibitem[\protect\citeauthoryear{{Li}, {Mo}  \& {Gao}}{{Li}
  et~al.}{2008}]{li2008}
{Li} Y.,  {Mo} H.~J.,   {Gao} L.,  2008, \mn@doi [\mnras]
  {10.1111/j.1365-2966.2008.13667.x}, \href
  {https://ui.adsabs.harvard.edu/abs/2008MNRAS.389.1419L} {389, 1419}

\bibitem[\protect\citeauthoryear{{Ludlow} et~al.,}{{Ludlow}
  et~al.}{2013}]{ludlow13}
{Ludlow} A.~D.,  et~al., 2013, \mn@doi [\mnras] {10.1093/mnras/stt526}, \href
  {https://ui.adsabs.harvard.edu/abs/2013MNRAS.432.1103L} {432, 1103}

\bibitem[\protect\citeauthoryear{{Mao}, {Zentner}  \& {Wechsler}}{{Mao}
  et~al.}{2018}]{conditional_binning}
{Mao} Y.-Y.,  {Zentner} A.~R.,   {Wechsler} R.~H.,  2018, \mn@doi [\mnras]
  {10.1093/mnras/stx3111}, \href
  {https://ui.adsabs.harvard.edu/abs/2018MNRAS.474.5143M} {474, 5143}

\bibitem[\protect\citeauthoryear{{McClintock} et~al.,}{{McClintock}
  et~al.}{2019}]{McClintock18}
{McClintock} T.,  et~al., 2019, \mn@doi [\mnras] {10.1093/mnras/sty2711}, \href
  {https://ui.adsabs.harvard.edu/abs/2019MNRAS.482.1352M} {482, 1352}

\bibitem[\protect\citeauthoryear{{Melchior} et~al.,}{{Melchior}
  et~al.}{2017}]{des_miscentering}
{Melchior} P.,  et~al., 2017, \mn@doi [\mnras] {10.1093/mnras/stx1053}, \href
  {https://ui.adsabs.harvard.edu/abs/2017MNRAS.469.4899M} {469, 4899}

\bibitem[\protect\citeauthoryear{{More}, {Diemer}  \& {Kravtsov}}{{More}
  et~al.}{2015}]{More2015TheMass}
{More} S.,  {Diemer} B.,   {Kravtsov} A.~V.,  2015, \mn@doi [\apj]
  {10.1088/0004-637X/810/1/36}, \href
  {https://ui.adsabs.harvard.edu/abs/2015ApJ...810...36M} {810, 36}

\bibitem[\protect\citeauthoryear{{More} et~al.,}{{More}
  et~al.}{2016}]{More2016DetectionClusters}
{More} S.,  et~al., 2016, \mn@doi [\apj] {10.3847/0004-637X/825/1/39}, \href
  {https://ui.adsabs.harvard.edu/abs/2016ApJ...825...39M} {825, 39}

\bibitem[\protect\citeauthoryear{{Moster}, {Naab}  \& {White}}{{Moster}
  et~al.}{2018}]{emerge}
{Moster} B.~P.,  {Naab} T.,   {White} S. D.~M.,  2018, \mn@doi [\mnras]
  {10.1093/mnras/sty655}, \href
  {https://ui.adsabs.harvard.edu/abs/2018MNRAS.477.1822M} {477, 1822}

\bibitem[\protect\citeauthoryear{{Moster}, {Naab}, {Lindstr{\"o}m}  \&
  {O'Leary}}{{Moster} et~al.}{2020}]{galaxynet}
{Moster} B.~P.,  {Naab} T.,  {Lindstr{\"o}m} M.,   {O'Leary} J.~A.,  2020,
  arXiv e-prints, \href {https://ui.adsabs.harvard.edu/abs/2020arXiv200512276M}
  {p. arXiv:2005.12276}

\bibitem[\protect\citeauthoryear{{Navarro}, {Frenk}  \& {White}}{{Navarro}
  et~al.}{1996}]{Navarro1996TheHalos}
{Navarro} J.~F.,  {Frenk} C.~S.,   {White} S. D.~M.,  1996, \mn@doi [\apj]
  {10.1086/177173}, \href
  {https://ui.adsabs.harvard.edu/abs/1996ApJ...462..563N} {462, 563}

\bibitem[\protect\citeauthoryear{{Nelson}, {Lau}  \& {Nagai}}{{Nelson}
  et~al.}{2014}]{neslon14_nthermal_pressure}
{Nelson} K.,  {Lau} E.~T.,   {Nagai} D.,  2014, \mn@doi [\apj]
  {10.1088/0004-637X/792/1/25}, \href
  {https://ui.adsabs.harvard.edu/abs/2014ApJ...792...25N} {792, 25}

\bibitem[\protect\citeauthoryear{{Nishimichi} et~al.,}{{Nishimichi}
  et~al.}{2021}]{Nishimichi21}
{Nishimichi} T.,  et~al., 2021, {DarkEmulator: Cosmological emulation code for
  halo clustering statistics} (\mn@eprint {ascl} {2103.009})

\bibitem[\protect\citeauthoryear{{O'Donnell}, {Behroozi}  \&
  {More}}{{O'Donnell} et~al.}{2020}]{odonnell2020}
{O'Donnell} C.,  {Behroozi} P.,   {More} S.,  2020, arXiv e-prints, \href
  {https://ui.adsabs.harvard.edu/abs/2020arXiv200508995O} {p. arXiv:2005.08995}

\bibitem[\protect\citeauthoryear{{Planck Collaboration} et~al.,}{{Planck
  Collaboration} et~al.}{2014}]{Ade2014PlanckParameters}
{Planck Collaboration} et~al., 2014, \mn@doi [\aap]
  {10.1051/0004-6361/201423743}, \href
  {https://ui.adsabs.harvard.edu/abs/2014A&A...571A..31P} {571, A31}

\bibitem[\protect\citeauthoryear{{Prada}, {Klypin}, {Cuesta}, {Betancort-Rijo}
  \& {Primack}}{{Prada} et~al.}{2012}]{Prada2011HaloCosmology}
{Prada} F.,  {Klypin} A.~A.,  {Cuesta} A.~J.,  {Betancort-Rijo} J.~E.,
  {Primack} J.,  2012, \mn@doi [\mnras] {10.1111/j.1365-2966.2012.21007.x},
  \href {https://ui.adsabs.harvard.edu/abs/2012MNRAS.423.3018P} {423, 3018}

\bibitem[\protect\citeauthoryear{{Rozo} \& {Rykoff}}{{Rozo} \&
  {Rykoff}}{2014}]{redmapper2}
{Rozo} E.,  {Rykoff} E.~S.,  2014, \mn@doi [\apj] {10.1088/0004-637X/783/2/80},
  \href {https://ui.adsabs.harvard.edu/abs/2014ApJ...783...80R} {783, 80}

\bibitem[\protect\citeauthoryear{{Rozo}, {Rykoff}, {Bartlett}  \&
  {Evrard}}{{Rozo} et~al.}{2014}]{redmapper1}
{Rozo} E.,  {Rykoff} E.~S.,  {Bartlett} J.~G.,   {Evrard} A.,  2014, \mn@doi
  [\mnras] {10.1093/mnras/stt2091}, \href
  {https://ui.adsabs.harvard.edu/abs/2014MNRAS.438...49R} {438, 49}

\bibitem[\protect\citeauthoryear{{Rozo}, {Rykoff}, {Bartlett}  \&
  {Melin}}{{Rozo} et~al.}{2015}]{redmapper3}
{Rozo} E.,  {Rykoff} E.~S.,  {Bartlett} J.~G.,   {Melin} J.-B.,  2015, \mn@doi
  [\mnras] {10.1093/mnras/stv605}, \href
  {https://ui.adsabs.harvard.edu/abs/2015MNRAS.450..592R} {450, 592}

\bibitem[\protect\citeauthoryear{{Rykoff} et~al.,}{{Rykoff}
  et~al.}{2016}]{rykoff16}
{Rykoff} E.~S.,  et~al., 2016, \mn@doi [\apjs] {10.3847/0067-0049/224/1/1},
  \href {https://ui.adsabs.harvard.edu/abs/2016ApJS..224....1R} {224, 1}

\bibitem[\protect\citeauthoryear{{Shi}, {Komatsu}, {Nelson}  \& {Nagai}}{{Shi}
  et~al.}{2015}]{shi15_nthermal_pressure}
{Shi} X.,  {Komatsu} E.,  {Nelson} K.,   {Nagai} D.,  2015, \mn@doi [\mnras]
  {10.1093/mnras/stv036}, \href
  {https://ui.adsabs.harvard.edu/abs/2015MNRAS.448.1020S} {448, 1020}

\bibitem[\protect\citeauthoryear{{Singh}, {Mandelbaum}, {Seljak}, {Slosar}  \&
  {Vazquez Gonzalez}}{{Singh} et~al.}{2017}]{Singh2017Galaxy-galaxyProperties}
{Singh} S.,  {Mandelbaum} R.,  {Seljak} U.,  {Slosar} A.,   {Vazquez Gonzalez}
  J.,  2017, \mn@doi [\mnras] {10.1093/mnras/stx1828}, \href
  {https://ui.adsabs.harvard.edu/abs/2017MNRAS.471.3827S} {471, 3827}

\bibitem[\protect\citeauthoryear{{Singh}, {Alam}, {Mandelbaum}, {Seljak},
  {Rodriguez-Torres}  \& {Ho}}{{Singh} et~al.}{2019}]{singh19}
{Singh} S.,  {Alam} S.,  {Mandelbaum} R.,  {Seljak} U.,  {Rodriguez-Torres} S.,
    {Ho} S.,  2019, \mn@doi [\mnras] {10.1093/mnras/sty2681}, \href
  {https://ui.adsabs.harvard.edu/abs/2019MNRAS.482..785S} {482, 785}

\bibitem[\protect\citeauthoryear{{Springel} et~al.,}{{Springel}
  et~al.}{2018}]{tng_illustris}
{Springel} V.,  et~al., 2018, \mn@doi [\mnras] {10.1093/mnras/stx3304}, \href
  {https://ui.adsabs.harvard.edu/abs/2018MNRAS.475..676S} {475, 676}

\bibitem[\protect\citeauthoryear{{Sunayama}, {Hearin}, {Padmanabhan}  \&
  {Leauthaud}}{{Sunayama} et~al.}{2016}]{Sunayama_scale_dependence}
{Sunayama} T.,  {Hearin} A.~P.,  {Padmanabhan} N.,   {Leauthaud} A.,  2016,
  \mn@doi [\mnras] {10.1093/mnras/stw332}, \href
  {https://ui.adsabs.harvard.edu/abs/2016MNRAS.458.1510S} {458, 1510}

\bibitem[\protect\citeauthoryear{{The LSST Dark Energy Science Collaboration}
  et~al.,}{{The LSST Dark Energy Science Collaboration} et~al.}{2018}]{lsst}
{The LSST Dark Energy Science Collaboration} et~al., 2018, arXiv e-prints,
  \href {https://ui.adsabs.harvard.edu/abs/2018arXiv180901669T} {p.
  arXiv:1809.01669}

\bibitem[\protect\citeauthoryear{{Wechsler}, {Bullock}, {Primack}, {Kravtsov}
  \& {Dekel}}{{Wechsler} et~al.}{2001}]{wechsler2001}
{Wechsler} R.~H.,  {Bullock} J.~S.,  {Primack} J.~R.,  {Kravtsov} A.~V.,
  {Dekel} A.,  2001, arXiv e-prints, \href
  {https://ui.adsabs.harvard.edu/abs/2001astro.ph.11069W} {pp
  astro--ph/0111069}

\bibitem[\protect\citeauthoryear{{Wechsler}, {Bullock}, {Primack}, {Kravtsov}
  \& {Dekel}}{{Wechsler} et~al.}{2002}]{wechsler02}
{Wechsler} R.~H.,  {Bullock} J.~S.,  {Primack} J.~R.,  {Kravtsov} A.~V.,
  {Dekel} A.,  2002, \mn@doi [\apj] {10.1086/338765}, \href
  {https://ui.adsabs.harvard.edu/abs/2002ApJ...568...52W} {568, 52}

\bibitem[\protect\citeauthoryear{{Wechsler}, {Zentner}, {Bullock}, {Kravtsov}
  \& {Allgood}}{{Wechsler} et~al.}{2006}]{Wechsler2006}
{Wechsler} R.~H.,  {Zentner} A.~R.,  {Bullock} J.~S.,  {Kravtsov} A.~V.,
  {Allgood} B.,  2006, \mn@doi [\apj] {10.1086/507120}, \href
  {https://ui.adsabs.harvard.edu/abs/2006ApJ...652...71W} {652, 71}

\bibitem[\protect\citeauthoryear{{Wetzel} \& {Nagai}}{{Wetzel} \&
  {Nagai}}{2015}]{wetzel_nagai15}
{Wetzel} A.~R.,  {Nagai} D.,  2015, \mn@doi [\apj]
  {10.1088/0004-637X/808/1/40}, \href
  {https://ui.adsabs.harvard.edu/abs/2015ApJ...808...40W} {808, 40}

\bibitem[\protect\citeauthoryear{{White} \& {Rees}}{{White} \&
  {Rees}}{1978}]{whiterees78}
{White} S.~D.~M.,  {Rees} M.~J.,  1978, \mn@doi [\mnras]
  {10.1093/mnras/183.3.341}, \href
  {https://ui.adsabs.harvard.edu/abs/1978MNRAS.183..341W} {183, 341}

\bibitem[\protect\citeauthoryear{{Wright} \& {Brainerd}}{{Wright} \&
  {Brainerd}}{2000}]{wright_brainerd_00}
{Wright} C.~O.,  {Brainerd} T.~G.,  2000, \mn@doi [\apj] {10.1086/308744},
  \href {https://ui.adsabs.harvard.edu/abs/2000ApJ...534...34W} {534, 34}

\bibitem[\protect\citeauthoryear{{Xhakaj}, {Diemer}, {Leauthaud}, {Wasserman},
  {Huang}, {Luo}, {Adhikari}  \& {Singh}}{{Xhakaj} et~al.}{2019a}]{xhakaj19}
{Xhakaj} E.,  {Diemer} B.,  {Leauthaud} A.,  {Wasserman} A.,  {Huang} S.,
  {Luo} Y.,  {Adhikari} S.,   {Singh} S.,  2019a, arXiv e-prints, \href
  {https://ui.adsabs.harvard.edu/abs/2019arXiv191109295X} {p. arXiv:1911.09295}

\bibitem[\protect\citeauthoryear{Xhakaj, Leauthaud, Diemer  \& Behroozi}{Xhakaj
  et~al.}{2019b}]{Xhakaj_2019}
Xhakaj E.,  Leauthaud A.,  Diemer B.,   Behroozi P.,  2019b, \mn@doi [Research
  Notes of the {AAS}] {10.3847/2515-5172/ab5579}, 3, 169

\bibitem[\protect\citeauthoryear{{van den Bosch} \& {Ogiya}}{{van den Bosch} \&
  {Ogiya}}{2018}]{numericalartifacts}
{van den Bosch} F.~C.,  {Ogiya} G.,  2018, \mn@doi [\mnras]
  {10.1093/mnras/sty084}, \href
  {https://ui.adsabs.harvard.edu/abs/2018MNRAS.475.4066V} {475, 4066}

\makeatother
\end{thebibliography}

\bsp	
\label{lastpage}
\end{document}